\documentclass[amsmath,amssymb,nofootinbib,superscriptaddress,showkeys,twocolumn]{revtex4-2}
\usepackage{graphicx}
\usepackage{dcolumn}
\usepackage{bm}
\usepackage{subfigure}
 \usepackage{multirow} 
\usepackage[colorlinks=true, citecolor=blue, urlcolor = magenta, linkcolor= red, bookmarks=true]{hyperref}
\usepackage{orcidlink}
\usepackage{booktabs} 

\begin{document}
\title{Probing Loop Quantum Gravity black holes through gravitational lensing}

\author{Arun Kumar\texorpdfstring{\href{https://orcid.org/0000-0001-8461-5368}{\orcidlink{0000-0001-8461-5368}}{}}}\email{arunbidhan@gmail.com}
\affiliation{Institute for Theoretical Physics and Cosmology, Zhejiang University of Technology, Hangzhou 310023, China}

\author{Qiang Wu\texorpdfstring{\href{https://orcid.org/0000-0002-5483-4903}{\orcidlink{0000-0002-5483-4903}}{}}} \email{wuq@zjut.edu.cn} 
\affiliation{Institute for Theoretical Physics and Cosmology, Zhejiang University of Technology, Hangzhou 310023, China}

\author{Tao Zhu\texorpdfstring{\href{https://orcid.org/0000-0003-2286-9009}{\orcidlink{0000-0003-2286-9009}}{}}} \email{zhut05@zjut.edu.cn} 
\affiliation{Institute for Theoretical Physics and Cosmology, Zhejiang University of Technology, Hangzhou 310023, China}

\author{Sushant~G.~Ghosh \texorpdfstring{\href{https://orcid.org/0000-0002-0835-3690}{\orcidlink{0000-0002-0835-3690}}{}}}\email{sghosh2@jmi.ac.in}
\affiliation{Centre for Theoretical Physics, Jamia Millia Islamia, New Delhi 110025, India}
\affiliation{Astrophysics and Cosmology Research Unit, 
School of Mathematics, Statistics and Computer Science, University of KwaZulu-Natal, Private Bag 54001, Durban 4000, South Africa}
	
\begin{abstract}

We investigate strong gravitational lensing by a charged loop quantum gravity (LQG) black hole obtained through the polymerisation scheme of Borges \textit{et al.} \cite{Borges:2023fog}. These effective geometries replace the Reissner--Nordstr\"om singularity with a symmetric transition surface and admit an extremal, cold remnant determined by the minimal area gap in LQG. In turn, we derive the null geodesic equations, investigate the photon effective potential, and obtain expressions for the photon-sphere radius and critical impact parameter. We compute the weak-field deflection angle and Einstein ring size, highlighting the deviations induced by the polymerisation parameter and the Barbero--Immirzi parameter. In the strong-field regime, we compute the strong deflection coefficients $(\bar{a},\bar{b})$ and evaluate the lensing observables $\theta_\infty$, $s$, and $r_{\rm mag}$. Unlike the Reissner--Nordstr\"om case, the LQG corrections enhance the deflection angle and increase the angular separation of relativistic images, with deviations growing as the geometry approaches the LQG remnant limit. We further compute the corresponding observables for Sgr~A* and M87*, finding that the quantum-gravity modifications lie within the potential sensitivity of next-generation VLBI facilities. For M87*, the angular separation $s\in(0.05712,0.19123)\,\mu\text{as}$, while it is $s\in(0.07595,0.25426)\,\mu\text{as}$ for Sgr A*. The relative flux ratio is found to lie in the range, $r_{\rm mag}\in(4.49272,5.96397)$. Our analysis demonstrates that LQG-induced corrections leave characteristic strong and weak-lensing imprints, offering a promising observational pathway to probe quantum gravity using near-future high-resolution observations.

\end{abstract}
	
	
\maketitle
	
\section{Introduction}\label{sec1}

Black holes (BHs) are among the most notable predictions of General Relativity (GR), which has been incredibly successful in explaining gravitation at macroscopic scales. Observational evidence from gravitational-wave detections by the LIGO–Virgo–KAGRA Collaboration \cite{LIGOScientific:2016aoc, LIGOScientific:2021djp} and horizon-scale imaging of supermassive BHs by the Event Horizon Telescope (EHT) \cite{EventHorizonTelescope:2019dse, EventHorizonTelescope:2022wkp} has opened new windows into strong-field gravity. The investigation of quantum-gravitational corrections to classical BH spacetimes and their possible detectable signature has been restored by these findings.
Gravitational lensing, a cornerstone prediction of general relativity, presents exceptional opportunities to investigate fundamental aspects of gravity and the structure of spacetime. While weak-field lensing offers important insights into the distribution of dark matter, the production of large-scale structures, and precise testing of general relativity \cite{Bartelmann:2010fz, Weinberg:1972kfs, Schneider:1992bmb}. In this regime, the deflection of light is small and can be treated perturbatively, making weak lensing highly sensitive to even subtle deviations from classical black hole geometries. The weak-lensing signatures of quantum-gravity-inspired black holes may provide a natural way to investigate quantum changes in the low-curvature, observably accessible regime.

When light grazes the photon sphere of BHs, it reveals intricate details of spacetime geometry in gravity's most extreme environments \cite{Virbhadra:1999nm, Virbhadra:2008ws}. The geometric framework of photon spheres established essential foundations for understanding these phenomena \cite{Claudel:2000yi}.  Strong gravitational lensing provides a straightforward means of distinguishing between classical GR BHs and quantum-corrected geometries, making it an ideal probe of the near-horizon area.  The general formalism for strong deflection lensing was established by Darwin \cite{1959RSPSA.249..180D} and later cast into an analytical framework by Bozza \cite{Bozza:2002zj}. Bozza's seminal work \cite{Bozza:2001xd} established a powerful analytical framework for strong-field gravitational lensing that has since been applied across diverse BH spacetimes. Subsequent extensions incorporated time delays as distance estimators \cite{Bozza:2003cp}, gravitational wave lensing \cite{Seto:2003iw}, and detailed analysis of various BH solutions, including the Kerr metric \cite{Vazquez:2003zm, Bozza:2005tg, Bozza:2006nm}. Research has progressively expanded beyond conventional general relativity solutions to examine exotic compact objects and modified gravity scenarios, including dilaton-axion BHs \cite{Gyulchev:2006zg, Gyulchev:2007zz}, naked singularities \cite{Virbhadra:2002ju, Atamurotov:2022srw}, Kiselev BHs \cite{Younas:2015sva, Azreg-Ainou:2017obt}, Einstein-\AE{}ther BH \cite{Zhu:2019ura}, and BHs in Eddington-inspired Born-Infeld gravity \cite{Wei:2014dka, Sotani:2015ewa, Babar:2021nst}. The emergence of 4D Einstein-Gauss-Bonnet (EGB) gravity has generated considerable interest due to its distinctive lensing signatures \cite{Islam:2020xmy, Kumar:2020sag, Islam:2022ybr}. Investigations have broadened to include BHs in Horndeski gravity \cite{Kumar:2021cyl, Walia:2021emv}, Simpson-Visser spacetimes \cite{Islam:2021ful, Ghosh:2023usx}, magnetized BHs \cite{Vachher:2025jsq}, and nonsingular alternatives such as Bardeen and Kerr-Sen BHs \cite{Ghosh:2020spb, Islam:2022ybr}. Plasma effects on lensing observables have emerged as another significant research direction \cite{Liu:2016eju, Bisnovatyi-Kogan:2017kii, Babar:2021nst, Feleppa:2024vdk, Turakhonov:2025ojy, Umarov:2025btg}. Particularly intriguing is the established connection between BH quasinormal modes and strong deflection lensing coefficients \cite{Stefanov:2010xz, Raffaelli:2014ola}.

Advances in understanding image distortions \cite{Virbhadra:2022iiy, Virbhadra:2024xpk} and the effects of the cosmological constant \cite{Adler:2022qtb} have refined our interpretation of observational signatures. The EHT observations have dramatically accelerated this field, providing direct constraints on BH metrics through shadow analysis and strong lensing features \cite{Kuang:2022xjp, Kuang:2022ojj, Kumar:2023jgh, Molla:2024yde, Zhao:2024hep, Vachher:2024ait}. These developments enable rigorous testing of supermassive compact objects with regular spacetimes \cite{Kumar:2022fqo}, string-inspired metrics \cite{Molla:2024lpt, Vachher:2024ezs, Ali:2024mrt}, flat ${\cal R}^2$ spacetime \cite{Yan:2025mlg}, hair Horndeski BH \cite{Shi:2024bpm}, and dark matter halo influences \cite{Vachher:2024ldc, Ali:2025rop, Ali:2025ney, Jusufi:2019nrn, Jusufi:2020cpn}.

Among various approaches to quantum gravity, LQG offers a non-perturbative and background-independent quantization of geometry, in which geometric operators possess discrete spectra. When applied to spherically symmetric spacetimes, LQG-inspired effective geometries lead to \emph{loop quantum BHs} (\emph{LQGBHs}), which resolve the central singularity and exhibit a non-singular core \cite{Modesto:2004wm, Modesto:2005zm, Modesto:2008im, Ashtekar:2018lag, Ashtekar:2020ckv}. Most of these constructions predict the existence of a minimal area scale, determined by the LQG area gap, leading to polymer-corrected effective metrics that modify the interior geometry while asymptotically recovering the classical Schwarzschild or RN exterior. 

A particularly compelling feature of LQBH models is the possible emergence of \emph{Planck-scale remnants}. In several quantization schemes—such as self-dual BHs \cite{Modesto:2008im}, polymerized geometries \cite{Bojowald:2005qw, Chiou:2008eg}, and quantum-extended Kruskal spacetimes \cite{Ashtekar:2020ckv}—the Hawking temperature reaches a maximum and subsequently drops to zero as the mass approaches a critical value, leading to an extremal, cold remnant. Phenomenological studies of several of these quantum corrected BH have been explored by several authors of this paper, see refs.~\cite{Jiang:2023img, Liu:2023vfh, Tu:2023xab, Liu:2020ola, Zhu:2020tcf, Yan:2022fkr, Yan:2023vdg, Jiang:2024vgn, Jiang:2024cpe, Uktamov:2024ckf, Xamidov:2024xpc, Xamidov:2025oqx} and references therein. A new class of charged polymerized BH model inside a single-parameter effective LQG framework was recently proposed by Borges \textit{et al.} \cite{Borges:2023fub}. BHs with an extremal configuration and vanishing surface gravity are produced by their model, which includes a transition surface defined by the smallest LQG area. In this model, as Hawking evaporation continues, the system asymptotically approaches a stable residual state, providing a theoretically motivated candidate for Planck-scale relics and dark matter.

LQG has recently gained prominence as a leading approach to quantum gravity, with its semi-classical implications for BH structure representing an active research frontier. LQG-inspired BHs typically exhibit novel characteristics, particularly the absence of central singularities, which produce distinctive imprints on strong lensing observables \cite{Kumar:2023jgh}. Recent investigations have begun exploring lensing properties of both static \cite{Wang:2024iwt} and rotating \cite{Kumar:2023jgh, Vachher:2024ait} BHs in quantum-gravity-motivated scenarios, underscoring their potential for observational testing. Examining lensing by these quantum-corrected BHs, including charged configurations, proves crucial for connecting quantum gravity predictions with astrophysical observations.

We observe that the gravitational lensing properties of the charged loop quantum gravity black hole (\emph{LQGBH}) proposed in Ref.~\cite{Borges:2023fub} remain largely unexplored. Due to the presence of an extremal remnant structure, a polymer-corrected interior, and altered relations between mass, charge, and the underlying quantum parameters, these charged \emph{LQGBHs} are very different from their classical RN counterparts. In this paper, we examine the weak and strong gravitational lensing of the \emph{LQGBHs} \cite{Borges:2023fub}, emphasising the possible observable footprints of quantum corrections caused by polymerisation, and compare the results with their GR counterparts. We compute the deflection angle in both the weak and strong-field limits and evaluate the associated observables that are crucial for compact-object lensing. We further highlight the imprints of quantum-gravity effects on the lensing signatures by examining how the lensing coefficients behave as the geometry approaches the remnant regime. In addition to offering possible restrictions on the polymerization parameters from future high-precision observations, our findings shed fresh light on the phenomenology of charged \emph{LQGBHs}. 

This paper is organised as follows: In  Sec. \ref{sec2}, we review the charged \emph{LQGBHs} \cite{Borges:2023fub}, summarising the key geometric features, the polymerisation scheme, and the constraints arising from the minimal area condition.  We discuss gravitational lensing in this LQG spacetime in Sec. \ref{sec3}. We first derive the photon geodesics and analyse both the weak-field and strong-field deflection angles, highlighting their deviations from the classical RN geometry. In Sec. \ref{3A}, we obtain the weak-deflection limit and the Einstein ring radius. Sec. \ref{3B} is devoted to the strong-field regime, where we compute the strong lensing coefficients and lensing observables using the Bozza--Tsukamoto formalism \cite{Bozza:2002zj, Tsukamoto:2016jzh}. In Sec. \ref{3C}, we apply our results to the supermassive BHs Sgr A* and M87*, estimating the lensing observables and comparing the predictions of charged \emph{LQGBHs} with those of GR. Finally, Sec. \ref{sec4} summarises our findings and discusses the astrophysical implications of strong lensing by \emph{LQGBHs} as potential probes of quantum gravity.

\section{Charged Loop Quantum Gravity BHs}\label{sec2}

Recent years have witnessed significant progress in developing LQG-motivated BH models, whose quantum corrections eliminate the classical singularity and may introduce deviations in the exterior geometry. Such deviations can influence accretion dynamics, gravitational wave signals, and, notably, strong gravitational lensing \cite{Bambi:2019tjh, Cunha:2018acu, Psaltis:2018xkc}. Consequently, strong lensing by LQG-corrected BHs offers a promising avenue to test quantum gravity effects in the strong-field regime \cite{Vagnozzi:2022moj, Afrin:2022ztr, Ghosh:2022mka}. Recently, \citet{Borges:2023fub} derived LQG-inspired charged regular spherically symmetric BH solutions using the polymerisation scheme \cite{ElizagaNavascues:2022rof, Alonso-Bardaji:2021yls}. In the polymer-based regular quantum BH models, the classical singularity is replaced by a symmetric transition between a BH and a white hole with the same mass. In this framework, the interior spacetime is described by the Ashtekar-Pawlowski-Singh (APS) model, which uses quantum-corrected Friedmann dynamics from loop quantum cosmology. Let us start with the classical Hamiltonian for RN, which, in terms of the canonical variables given in \cite{Modesto:2008im, Ashtekar:2018lag}, reads as \cite{Alonso-Bardaji:2023niu, Tibrewala:2012xb, Gambini:2014qta}
\begin{equation}\label{HM1}
    H=-\frac{1}{2G\gamma }\Big[p_b\left( b+\frac{\gamma^2}{ b}-\frac{\gamma^2 Q^2}{ bp_c}\right)+2cp_c\Big],
\end{equation} 
with $c$, $p_c$, and $b$, $p_b$ being conjugate variable pairs such that $\{c,p_c\}=2G\gamma$ and $\{b,p_b\}=G\gamma$. Using the polymerisation scheme \cite{Gambini:2021uzf}, \citet{Borges:2023fub} transformed the canonical variables into quantum operators to get the following LQG Hamiltonian  
\begin{eqnarray}\label{HM2}
    H_{LQG}&=&-\frac{1}{2G\gamma \sqrt{1+\gamma^2\delta_b^2}}\Big[p_b\Big(\frac{\sin(\delta_b b)}{\delta_b}+\frac{\gamma^2\delta_b}{\sin(\delta_b b)}\nonumber\\&&-\frac{\gamma^2\delta_b Q^2}{\sin(\delta_b b)p_c}\Big)+2cp_c\cos(\delta_b b)\Big],
\end{eqnarray} 
where $\delta_b$ and $\gamma$ signify the polymerisation and the Barbero-Immirzi parameters, respectively. and  Solving the dynamical equations of the Hamiltonian \eqref{HM2} along with some suitable variable transformation leads to the asymptotically flat charged \emph{LQGBHs} BH metric \cite{Borges:2023fub}
\begin{equation}\label{metric}
ds^2= -A(r)dt^2 + B(r)dr^2+C(r)\left(d\theta^2+\sin^2\theta{d\phi^2}\right),
\end{equation} 
with 
\begin{eqnarray}
A(r)&=&1-\frac{2M}{r}+\frac{Q^2}{r^2},\nonumber\\~B(r)&=&\left[A(r)\left(1-\frac{r_0}{M}g(r)\right)\right]^{-1},\nonumber\\C(r)&=&r^2,
\end{eqnarray} 
where $M$ and $Q$ are the BH mass and electric charge, respectively, and the radius of the transition surface, $r_0$, and the function $g(r)$ reads \cite{Borges:2023fub}
\begin{eqnarray}
r_0&=&\frac{(b_0^2-1)M}{b_0^2}\left(1+\sqrt{1-\frac{b_0^2Q^2}{(b_0^2-1)^2M^2}}\right),\\
g(r)&=&\frac{2Mr-Q^2}{r^2\left(1+\sqrt{1-\frac{b_0^2Q^2}{(b_0^2-1)^2M^2}}\right)},
\end{eqnarray}
such that the electric charge, $Q$, follows the following constraint \cite{Borges:2023fub}
\begin{equation}\label{Qcon}
    |Q|\leq\frac{\sqrt{b_0^2-1}}{b_0}M~~~~~~\text{for}~~~~~~b_0>1.
\end{equation} 
The quantity, $b_0$, is related to the polymerisation parameter, $\delta_b$, and the Barbero-Immirzi parameter, $\gamma$ via $b_0=\sqrt{1+\gamma\delta_b^2}$. When the polarisation parameter, $\delta_b\to0$ or $b_0=1$, the solution (\ref{metric}) reduces to the well-known RN, charged BH solution of classical theory. The metric \eqref{metric} exhibits the same horizon radii as RN, i.e, 
\begin{equation}\label{horizon}
    r_{\pm}=M(1\pm\sqrt{1-Q^2/M^2}).
\end{equation} 
The minimal area in LQG denotes the smallest nonzero value inside the discrete spectrum of the area operator, embodying the concept that the space at the Planck scale is quantised and granular rather than continuous. We can correlate the polymerisation parameter $\delta_b$ with the constants of motion, i.e., $M$ and $Q$, by applying the minimal area condition \cite{Modesto:2008im}
\begin{equation}\label{area}
    4\pi r_0^2=4\pi\sqrt{3}\gamma, 
\end{equation}\label{marea} as
\begin{equation}\label{deltab}
    \delta_b^2=\frac{\sqrt{3}}{\gamma\left(2\times 3^{1/4}\sqrt{\gamma}M-Q^2-\sqrt{3}\gamma\right)}.
\end{equation}Following the polymerisation parameter relation (\ref{deltab}), we can filter the physically significant dynamical trajectories in the context of LQG ( which satisfy the minimal area condition of LQG \eqref{area}).

Next, we employ Eq. \eqref{deltab} along with Eq. \eqref{Qcon} to refine the constraint on the BH charge as
\begin{equation}\label{Qcon1}
   Q^2\leq  3^{1/4}\sqrt{\gamma}M.
    \end{equation} From Eq. \eqref{deltab} we have $2\times 3^{1/4}\sqrt{\gamma}M-Q^2-\sqrt{3}\gamma>0$, by using this condition along with Eq. \eqref{horizon}, we can show that $r_-<r_0<r_+$ or it can be concluded that during BH evaporation a transition from BH to white hole will take place at the surface of the radius $r_0$ before reaching the Cauchy horizon. Before moving forward, let us talk about the bounds on $\gamma$. By taking into account the Bekenstein-Hawking entropy, the authors in Ref. \cite{Domagala:2004jt} estimated  $\ln2\leq\pi\gamma\leq\ln3$, but \citet{Meissner:2004ju} calculated the exact value of $\gamma$ to be $\gamma\approx0.23753$. Then, the authors in \cite{Corichi:2006bs} showed that for $\gamma\approx0.274$, the entropy of large BHs in LQG perfectly satisfies the Bekenstein-Hawking entropy. However, \citet{Pigozzo:2020zft} suggested that $\gamma$ should be $\sqrt{3}/6$ instead of $0.274$ for the Bekenstein-Hawking entropy area law to be satisfied. 
    
\section{ Gravitational Lensing}\label{sec3}

The phenomenon of deviation in the geodesics of light rays as they propagate through a gravitational field is termed gravitational lensing. There are two regimes of gravitational lensing: when light travels very far away from the massive object (lens), it is referred to as the weak field limit, but when light propagates very close to the gravitational object where the gravitational field is so prominent that the angular deflection becomes divergent at a particular approach limit, it is known as the strong field limit. Let us derive the geodesic equations next, so that we can examine the two limits. 

The Lagrangian dictating the photon trajectories in the equatorial plane ($\theta=\pi/2$) of charged \emph{LQGBHs} reads 
\begin{eqnarray}\label{Lagrangian}
    \mathcal{L}&=&\left(1-\frac{2}{r}+\frac{Q^2}{r^2}\right)\left(\frac{dt}{d\lambda}\right)^2\nonumber\\&&+\frac{1}{\left(1-\frac{2}{r}+\frac{Q^2}{r^2}\right)\left(1-\frac{b_0^2}{(b_0^2-1)}\left(\frac{2}{r}-\frac{Q^2}{r^2}\right)\right)}\left(\frac{dr}{d\lambda}\right)^2\nonumber\\&&+r^2\left(\frac{d\phi}{d\lambda}\right)^2.
\end{eqnarray}
Here, $\lambda$ denotes the affine parameter, and $r$, $t$ are the rescaled quantities such that $r\to r/M$, $t\to t/M$, respectively, with $M=1$.
Using the Euler-Lagrange equations with respect to the Lagrangian \eqref{Lagrangian}, we obtained two conserved quantities corresponding to the Killing vectors associated with time translation $\partial_t$ and rotational invariance $\partial_\phi$ \cite{Chandrasekhar:1985kt} as
\begin{equation}\label{conserevQ}
L= r^2\frac{d\phi}{d\lambda},  \qquad \qquad E=-\left(1-\frac{2}{r}+\frac{Q^2}{r^2}\right)\frac{dt}{d\lambda}.
\end{equation}Hence, using the conserved quantities, Eq. (\ref{conserevQ}), in Eq. (\ref{Lagrangian}) and applying the null geodesics condition, $\mathcal{L}=0$, results in
\begin{eqnarray}\label{dr}
\left(\frac{dr}{d\lambda}\right)^2&=& \left(1-\frac{b_0^2}{(b_0^2-1)}\left(\frac{2}{r}-\frac{Q^2}{r^2}\right)\right)\nonumber\\&&\times\left[E^2- \frac{L^2}{r^2}\left(1-\frac{2}{r}+\frac{Q^2}{r^2}\right)\right].
\end{eqnarray}
We use the radial motion equation, $(dr/d\lambda)^2+V_{\text{eff}}=0$, to identify the effective potential as
\begin{eqnarray}
 \frac{V_{\text{eff}}}{E^2}&=& \left(1-\frac{b_0^2}{(b_0^2-1)}\left(\frac{2}{r}-\frac{Q^2}{r^2}\right)\right)\nonumber\\&&\times\left[-1+ \frac{u^2}{r^2}\left(1-\frac{2}{r}+\frac{Q^2}{r^2}\right)\right],
\end{eqnarray}where $u=L/E$ signifies the impact parameter. The photon orbits $r_{\text{ps}}$ and the critical impact parameter $u_{\text{ps}}$ for charged \emph{LQGBHs} could be found by using $V_{\text{eff}}|_{r=r_{\text{ps}}}=0\equiv dV_{\text{eff}}/dr|_{r=r_{\text{ps}}}$ and $d^2V_{\text{eff}}/dr^2|_{r=r_{\text{ps}}}<0$ as
\begin{equation}
    r_{\text{ps}}=\frac{1}{2}\left(3+\sqrt{9-8Q^2}\right),~~~~~~~~~u_{\text{ps}}=\frac{r_{\text{ps}}^2}{\sqrt{r_{\text{ps}}-Q^2}},
\end{equation} 
which are exactly the same as the expressions of $r_{\text{ps}}$ and $u_{\text{ps}}$ of RN BHs \cite{Bozza:2002zj,Tsukamoto:2016jzh,KUMAR2026170291}. Hence, we can conclude that $\delta_b$ and $\gamma$ do not affect the general form of these quantities, but it is important to mention that there has been a constraint on $Q$ given in Eq. \eqref{Qcon1} due to $\delta_b$ and $\gamma$. However, the allowed range of the charge $Q$ is significantly affected by the LQG corrections. As shown by the constraint \eqref{Qcon1}, quantum-gravity effects restrict the allowed parameter space. For instance, with $\gamma=\sqrt{3}/6$ and $M=1$, the maximum permissible charge is $Q_{\max}=0.840896$—smaller than the extremal limit $Q=1$ of the classical RN BHs. 

\subsection{Deflection Angle in Weak Gravitational Field Limits and Einstein Ring}\label{3A}

Gravitational lensing in the weak-field regime remains one of the most robust and extensively tested predictions of GR \cite{Weinberg:1972kfs, Bartelmann:2010fz}. In the weak-field limit, the deflection angle can be computed using a perturbative expansion of the metric functions, yielding the familiar post-Newtonian expressions that successfully describe lensing by stars, galaxies, and galaxy clusters. The \emph{LQGBHs} may induce small but potentially observable corrections to the standard weak-field bending of light \cite{Berti:2015itd, Cardoso:2019rvt}. Investigating these deviations is crucial, as high-precision astrometric surveys—such as VLTI/GRAVITY and future missions—can probe minute departures from GR predictions in the lensing observables. In this section, we derive the weak deflection angle for the charged \emph{LQGBHs} metric (\ref{metric}) and analyze the resulting modifications to the Einstein ring.  

\begin{figure}
\centering
	{\includegraphics[width=0.5\textwidth]{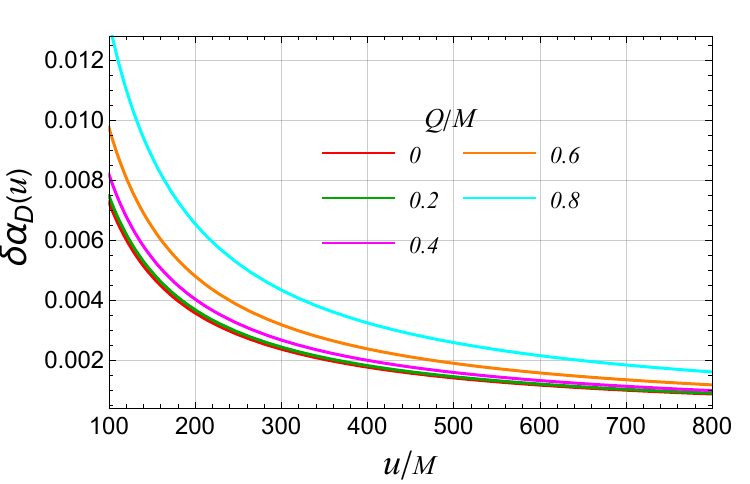}}
    \caption{The deviation in the weak--field deflection angle, 
$\delta\alpha_{D}(u) = \alpha_{D}(u) - \alpha_{D}(u)|_{\rm RN}$, 
for \emph{LQGBHs} relative to the RN  case. 
The Barbero--Immirzi parameter is fixed to $\gamma = \sqrt{3}/6$. 
The positive deviation increases with the charge $Q$, reflecting the enhancement of the deflection due to LQG polymerisation effects.}
\label{plot1}
\end{figure}

To set up the elements needed for calculating the deflection of light, consider a photon that originates in an asymptotically flat region and moves toward the BH, reaching a minimum radial distance $r_0$ from the center—the turning point—where $r_0>r_{\text{ps}}$. After being bent by the BH’s gravitational field, the photon continues to another asymptotically flat region. At this turning point, the condition $V_{\text{eff}}=0$ holds, leading to the following expression
\begin{equation}\label{ip}
    \frac{1}{u^2}=\frac{1}{r_0^2}\left(1-\frac{2}{r_0}+\frac{Q^2}{r_0^2}\right).
\end{equation}Now, by using Eqs. \eqref{conserevQ} and \eqref{dr}, we can express the angle of deflection for charged \emph{LQGBHs} as \cite{Bozza:2002zj}
\begin{eqnarray}\label{def1}
    \alpha_{\text{D}}(r_0)&=&2 \int_{r_0}^{\infty}\frac{d\phi}{dr} dr-\pi\nonumber\\&&\equiv 2 \int_{r_0}^{\infty}\Bigg[\left(1-\frac{b_0^2}{(b_0^2-1)}\left(\frac{2}{r}-\frac{Q^2}{r^2}\right)\right)\nonumber\\&&\left[\frac{r^4}{u^2}- r^2\left(1-\frac{2}{r}+\frac{Q^2}{r^2}\right)\right]\Bigg]^{-1/2}-\pi.
\end{eqnarray} 
Next, to obtain the deflection angle approximation in weak field limits, we proceed by defining a new variable, $\omega=r_0/r$, and taking the Taylor series expansion to get  
\begin{eqnarray}\label{def2}
  \alpha_{\text{D}}(r_0)&\approx&\int_{0}^{1}\frac{2}{\sqrt{1-\omega^2}}+\frac{2(1+\omega+\omega^2)}{r_0(1+\omega)\sqrt{1-\omega^2}}+\frac{\mathcal{A}}{b_0^4 r_0^2(1+\omega)^2}\nonumber\\&&+\frac{\mathcal{B}}{b_0^6 r_0^2 \sqrt{1-\omega^2 }(1+\omega)^3}-\pi,
\end{eqnarray} 
with
\begin{eqnarray}
    \mathcal{A}&=&3 \omega ^2 (1+\omega)^2+b_0^2 \omega  (1+\omega) \left( -2+\omega  (1+\omega)\left(Q^2-8\right)\right)\nonumber\\&&-b_0^4 \Bigg[-3-8 \omega (1+\omega) \left(1+\omega+\omega ^2\right)\nonumber\\&&+Q^2\left(1+2 \omega ^2\right)(1+\omega)^2 \Bigg]\nonumber\\
    \mathcal{B}&=&-5 \omega^3 (1+\omega)^3-b_0^2\omega^2(1+\omega)^2\left[-1+\omega(1+\omega)(Q^2-6)\right]\nonumber\\&&+b_0^4\omega(1+\omega)\Bigg[-3-12\omega(1+\omega)(1+2\omega(1+\omega))\nonumber\\&&+Q^2(1+\omega)(1+2\omega(1+2\omega)^2)\Bigg]\nonumber\\&&+b_0^6\Bigg[\left(1+2\omega(1+\omega)\right)\left(5+8\omega(1+\omega)(1+\omega+\omega^2)\right)\nonumber\\
&&-Q^2(1+\omega)^2\left(3+4\omega(1+2\omega(1+\omega+\omega^2))\right)\Bigg]\nonumber.
\end{eqnarray}
Integration of the integrand in Eq. \eqref{def2} leads to the deflection angle in terms of the distance of closest approach $r_0$ as 
\begin{eqnarray}\label{def3}
    &&\alpha_{\text{D}}(r_0)\simeq\frac{2(3b_0^2-1)}{b_0^2r_0}+\frac{1}{4b_0^4r_0^2}\Bigg[3\pi\nonumber\\&&+b_0^2\left(8+\pi(Q^2-12)\right)-4b_0^4\left(6+\pi(Q^2-6)\right)\Bigg]\nonumber\\&&+\frac{1}{b_0^6r_0^3}\Bigg[-\frac{10}{3}+b_0^2\left(18-\frac{3\pi}{2}-2Q^2\right)\nonumber\\&&+b_0^4\left(-41+6\pi+\frac{Q^2}{6}(50-3\pi)\right)\nonumber\\&&+b_0^6\left(67-12\pi+\left(2\pi-\frac{61}{3}\right)Q^2\right)\Bigg]+\mathcal{O}\left(\frac{1}{r_0^4}\right).
\end{eqnarray}
Next, we want to find the expression of the deflection angle in terms of the impact parameter $u$, for which we proceed with expanding $1/r_0$ in terms of $1/u$ as follows
\begin{equation}\label{ru}
    \frac{1}{r_0}\simeq\frac{1}{u}+\frac{1}{u^2}+\frac{5-Q^2}{2u^3}+\mathcal{O}\left(\frac{1}{u^4}\right).
\end{equation}
Inserting Eq. \eqref{ru} in Eq. \eqref{def3}, results in
\begin{eqnarray}\label{def4}
   && \alpha_{\text{D}}(b)\simeq\frac{2M}{b}\left[3-\frac{1}{b_0^2}\right]+\frac{\pi}{4b^2}\Bigg[3M^2\left(8-\frac{4}{b_0^2}+\frac{1}{b_0^4}\right)\nonumber\\&&+Q^2\left(-4+\frac{1}{b_0^4}\right)\Bigg]+\frac{2M}{3b^2}\Bigg[2M^2\left(105-\frac{63}{b_0^2}+\frac{27}{b_0^4}-\frac{1}{b_0^6}\right)\nonumber\\&&-2Q^2\left(\frac{35}{b_0^2}-\frac{14}{b_0^4}+\frac{3}{b_0^6}\right)\Bigg]+\mathcal{O}\left(\frac{1}{b^4}\right),
\end{eqnarray}
where $b=uM$ denotes the rescaled impact parameter. Taking the classical limits, i.e., inserting $b_0=1$ into Eq. \eqref{def4}, we recover the deflection angle of RN BHs, which reads
\begin{eqnarray}\label{defRN}
    \alpha_{\text{D}}(b)&\simeq&\frac{4M}{b}+\frac{3\pi}{4b^2}\left(5M^2-Q^2\right)\nonumber\\&&+\frac{M}{3b^3}\left(128M^2-48Q^2\right)+\mathcal{O}\left(\frac{1}{b^4}\right),
\end{eqnarray} 
which further reduces to the deflection angle of Schwarzschild BHs \cite{KumarWalia:2022ddq, Kumar:2025bim}, in the absence of electrodynamics ($Q=0$). We depicted how the deflection angle of charged \emph{LQGBHs} in weak field limits deviates from that of RN BHs in Fig. \ref{plot1}. Because the deviation is positive, we can conclude that the polymerisation has amplified the deflection. It can also be seen clearly from Fig. \ref{plot1} that the enhancement in deflection of charged \emph{LQGBHs} grows with electric charge $Q$. 

Next, we analyze how polymerisation affects the size of the Einstein Ring in charged BHs. We consider the case where the source, lens, and observer are exactly aligned, with both the source and observer in regions that are nearly flat. For our analysis, we use observational data from the Einstein ring of the galaxy ESO325-G004, which has a mass of $M = 1.50 \times 10^{11} M_{\odot}$ \cite{Smith:2005pq, Smith:2013ena}, where $M_{\odot} = 1.98 \times 10^{30}\, {\rm kg}$ is the solar mass. This mass includes the central BH as well as both dark and visible matter in the galaxy. The background source galaxy has a higher redshift of $z_s=1.141$. Hubble’s law relates the spectroscopic redshift to a galaxy’s proper distance with the equation $$cz = H_{0} D,$$ where $H_{0} = 70.4\, {\rm km}\, {\rm s}^{-1}\, {\rm Mpc}^{-1}$ is the Hubble constant and $D$ is the proper distance. In a flat universe, the comoving distance $d$ is given by $d = D(1+z)$. Using these equations, we find the distances between the lens, source, and observer.
\begin{align}
D_{\text{LS}}=\dfrac{c z_s(1+z_s)}{H_0}=2.863\times10^4\,{\rm Mpc}, \nonumber\\ D_\text{L}=\dfrac{c z_l(1+z_\ell)}{H_0}=1.542\times10^2\,{\rm Mpc} \ .\label{ESOdata}
\end{align} 

The angular size of the Einstein ring measured for the galaxy ESO325-G004 at redshift $z_l=0.0345$ is reported \cite{Smith:2005pq, Smith:2013ena}
\begin{equation}\label{obsering}
    \theta_{\text{E}}^{\rm obs} = (2.85^{+0.55}_{-0.25})'' .  
\end{equation}

\begin{figure}
\centering
	{\includegraphics[width=0.5\textwidth]{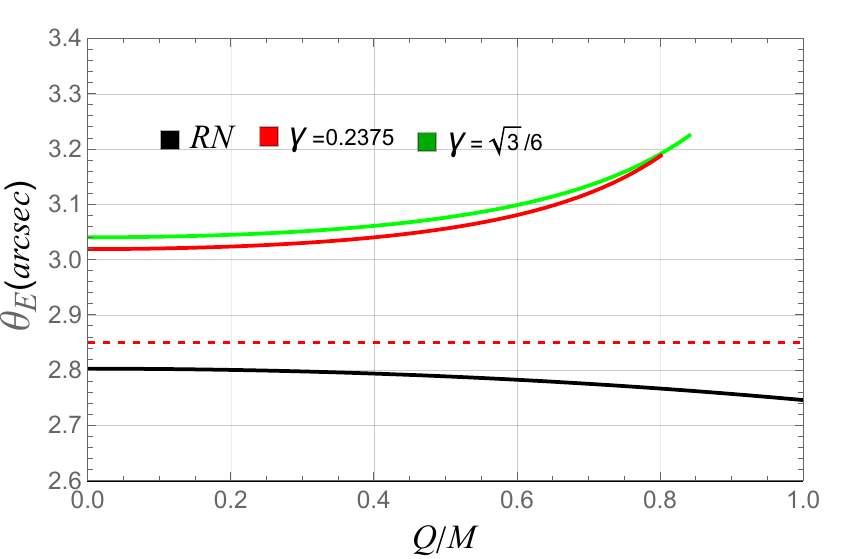}}
    \caption{The angular radius of the Einstein ring $\theta_{E}$ plotted as a function of the charge $Q$ for RN and charged \emph{LQGBHs}. 
The red dotted line marks the observed value $\theta^{\rm obs}_{E} = 2.85$ arcsec and the range of y-axis denoting the $1\sigma$ uncertainty. 
LQG corrections lead to an increase in $\theta_{E}$ with $Q$, in contrast to the decreasing trend in the RN case.}
\label{plot2}
\end{figure}

Bozza \cite{Bozza:2008ev} analysed various lens equations and found that the Ohanian lens equation and its associated forms are the most reliable approximations to the precise lens equation. In this context, they also developed a reformulation of the Ohanian lens equation in terms of the distances between the observer, lens, and source planes, which takes the form
\begin{align}\label{LE}
D_\text{S}\tan\mathcal{B}=\dfrac{D_\text{L}\sin\theta-D_\text{LS}\sin(\alpha_\text{D}-\theta)}{\cos(\alpha_\text{D}-\theta)},
\end{align}
where $\mathcal{B}$ denotes the angular position of the unlensed source, $\theta$  the angular position of an image, $D_\text{L}$ the distance from the observer to the lens plane, and $D_\text{LS}$ the distance from the lens to the source plane, with the relation $D_\text{S}=D_\text{LS}+D_\text{L}$ and $\sin\theta=b/D_\text{L}$. Under perfect alignment of the source, lens, and observer, the angular position of the source reduces to $\mathcal{B}=0$, leading to $\theta=\theta_\text{E}$, where $\theta_\text{E}$ represents the angular radius of the Einstein ring. In this limit, it becomes evident that $\alpha_\text{D}$ and $\theta$ are of order $\mathcal{O}(\epsilon)$ when $\mathcal{B}=0$. Therefore, for sufficiently small values of $\alpha_\text{D}$, $\theta$, and $\mathcal{B}=0$, Eq.~(\ref{LE}) can be simplified accordingly as
\begin{align}\label{Einsteinring}
\theta_\text{E}\simeq \frac{D_{\text{LS}}}{D_\text{S}}\alpha_\text{D},
\end{align} 
which, on substitution of the deflection angle \eqref{def4} along with the relation $\theta\simeq b/D_\text{L}$, takes the form 
\begin{eqnarray}\label{newER}
\theta_\text{E}&=&\dfrac{D_{\text{LS}}}{D_\text{S}}\Bigg[\frac{2M D_\text{L}}{\theta_\text{E}}\left(3-\frac{1}{b_0^2}\right)\nonumber\\&&+\frac{\pi M^2 D_\text{L}^2}{4\theta_\text{E}^2}\Bigg[3\left(8-\frac{4}{b_0^2}+\frac{1}{b_0^4}\right)+\frac{Q^2}{M^2}\left(-4+\frac{1}{b_0^4}\right)\Bigg]\nonumber\\&&+\frac{2M^3 D_\text{L}^3}{3\theta_\text{E}^3}\Bigg[2\left(105-\frac{63}{b_0^2}+\frac{27}{b_0^4}-\frac{1}{b_0^6}\right)\nonumber\\&&~~~~~~~-\frac{2Q^2}{M^2}\left(\frac{35}{b_0^2}-\frac{14}{b_0^4}+\frac{3}{b_0^6}\right)\Bigg]\Bigg]+ \mathcal{O}\left(\frac{M}{\theta_\text{E}}\right)^4.
\end{eqnarray} 
We solved Eq. \eqref{newER} numerically using the observational data of the galaxy ESO325-G004 to obtain $\theta_\text{E}$ and depicted those results in Fig. \ref{plot2}. The dotted red line in Fig. \ref{plot2} signifies $\theta_\text{E}=2.85$ arcsec while the light yellow region denotes the 1$\sigma$ uncertainties. Clearly, the angular size of the Einstein ring for charged \emph{LQGBHs} increases with increasing charge $Q$ and Barbero-Immirzi parameter $\gamma$ (red and green curves in Fig. \ref{plot2}), whereas for RN BHs (black curve) $\theta_{\text{E}}$ decreases with $Q$. We used widely accepted values of Barbero-Immirzi parameter, $\gamma=0.2375$ \cite{Meissner:2004ju} and $\gamma=\sqrt{3}/6$ \cite{Pigozzo:2020zft}. It is evident from Fig. \ref{plot2}, that charged \emph{LQGBHs} as well as RN BHs satisfy the observational data of the galaxy ESO325-G004 within 1$\sigma$ uncertainties for the entire parameter space. The deflection angle for charged \emph{LQGBHs} leads to small corrections to the Einstein ring radius, potentially measurable for astrophysical lenses with high-precision astrometry \cite{Bartelmann:2010fz, Schneider:1992bmb} (cf. Fig. \ref{plot2}). Thus, the weak-lensing properties of the charged \emph{LQGBHs}  provide an independent observational track—complementary to strong lensing and shadow measurements—for probing quantum-gravity–induced deviations from general relativity.

\subsection{Deflection Angle in Strong Gravitational Field limits and Observables}\label{3B}

Strong gravitational lensing is a powerful probe of the spacetime geometry in the vicinity of compact objects like BHs and has long been recognised as an important tool for testing gravity in the strong-field regime \cite{1959RSPSA.249..180D, Chandrasekhar:1985kt}. The strong deflection limit formalism \cite{Bozza:2002zj, Bozza:2001xd} has enabled precise analytic descriptions of relativistic images generated near the photon sphere of BHs. 
The LQG BHs are a natural arena for searching for such deviations, as their modified near-horizon and photon-sphere structures can leave observable imprints in strong lensing observables. Analyzing the strong lensing properties of this LQG-inspired regular BH is crucial for identifying potential observational signatures of quantum gravity. In what follows, we perform the strong gravitational lensing by a charged \emph{LQGBHs} (\ref{metric}) and calculate the lensing observables, highlighting the role of polymerisation-induced corrections in altering the strong-field lensing behaviour.

Now, we analyse the deflection angle and lensing coefficients for photons in the strong gravitational field regime ($r_0\to r_\text{ps}$) of the charged LQG spacetime, along with how its parameters affect these lensing coefficients in the strong-field regime. When we substitute the impact parameter \eqref{ip} in the deflection angle expression defined in Eq. \eqref{def1}, it takes the following form 
\begin{eqnarray}\label{sdef1}
    \alpha_{\text{D}}(r_0)&=& 2 \int_{r_0}^{\infty}\Bigg[\left(1-\frac{b_0^2}{(b_0^2-1)}\left(\frac{2}{r}-\frac{Q^2}{r^2}\right)\right)\nonumber\\&&\left[\frac{r^4}{r_0^2}\left(1-\frac{2}{r_0}+\frac{Q^2}{r_0^2}\right)- r^2\left(1-\frac{2}{r}+\frac{Q^2}{r^2}\right)\right]\Bigg]^{-1/2}\nonumber\\&&-\pi.
\end{eqnarray}
It is a well-established fact that the deflection angle \eqref{sdef1} exhibits divergence at the photon orbit radius $r_\text{ps}$, and can not be solved directly; we instead perform an expansion around $r_\text{ps}$ \cite{Virbhadra:1999nm,Claudel:2000yi,Bozza:2002zj}. By using Tsukamoto's recipe \cite{Tsukamoto:2016jzh}, which is an improved version of Bozza's method \cite{Bozza:2002zj}, we defined a  new variable $z=1-r_0/r_\text{ps}$ to obtain the analytical expression of the strong field deflection angle of the charged \emph{LQGBHs} in terms of impact parameter $u$, which reads   
\cite{Bozza:2002zj,Tsukamoto:2016jzh,Kumar:2020sag,Islam:2020xmy}
\begin{eqnarray}\label{sdef2}
\alpha_{D}(u) &=& \bar{a} \log\left(\frac{u}{u_{ps}} -1\right) + \bar{b} + \mathcal{O}(u-u_{ps}).   
\end{eqnarray}

\begin{figure}
\centering
	{\includegraphics[width=0.5\textwidth]{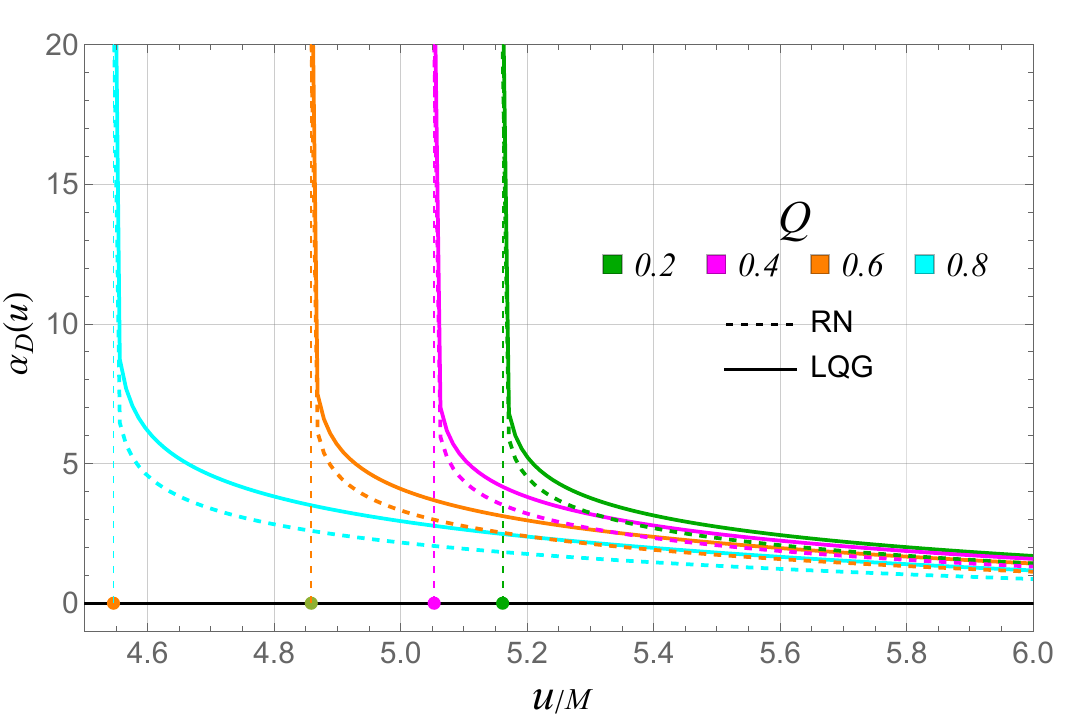}}
    \caption{The strong--field deflection angle $\alpha_{D}(u)$ as a function of the impact parameter $u$ for various values of the electric charge $Q$. 
Both RN (dashed curves) and charged \emph{LQGBHs} (solid curves) show divergence at $u = u_{\rm ps}$, but \emph{LQGBHs} exhibit consistently larger deflection for $u > u_{\rm ps}$. 
We fix the Barbero--Immirzi parameter to $\gamma = \sqrt{3}/6$.}
\label{plot3}
\end{figure}

\begin{figure*}
	\begin{centering}
		\begin{tabular}{c c}
		  \includegraphics[width=0.5\textwidth]{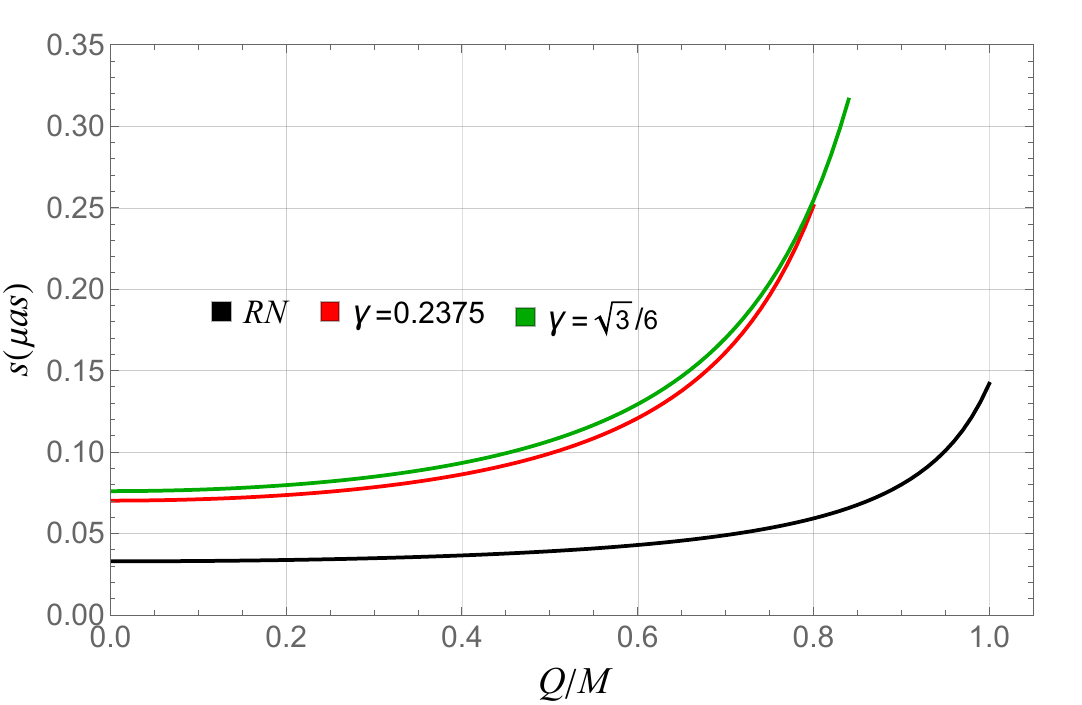}&
			\includegraphics[width=0.5\textwidth]{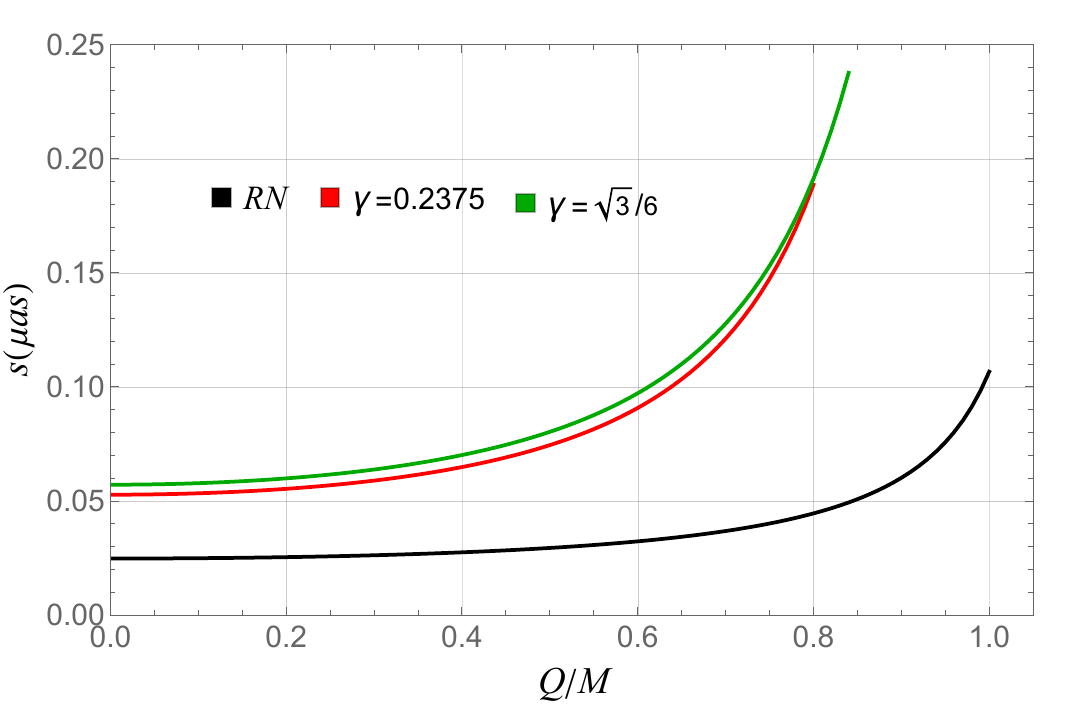}
			\end{tabular}
	\end{centering}
	\caption{The strong lensing observable $s = \theta_{1} - \theta_{\infty}$ is displayed as a function of the charge $Q$ for Sgr~A$^{*}$ (left) 
and M87$^{*}$ (right). Results are shown for RN and charged \emph{LQGBHs}, with $\gamma = 0.2375$ and $\gamma = \sqrt{3}/6$. LQG corrections enlarge the angular separation $s$, making relativistic images more widely separated than in the RN case.} \label{plot4}		
\end{figure*}

We obtained the analytical expression for the strong deflection coefficient $\bar{a}$, which is given as  
\begin{equation}
    \bar{a}=\frac{b_0^2r_\text{ps}^2}{\sqrt{\left(r_\text{ps}^2-2Q^2\right)\left[(b_0^2-1)Q^2+r_\text{ps}\left(2(1-b_0^2)+b_0^2r_\text{ps}\right)\right]}},
\end{equation}
which on taking $b_0=1$ and doing some basic calculations reduces to $\bar{a}$ for RN BHs \cite{Tsukamoto:2016jzh}  $$\bar{a}=\frac{r_\text{ps}}{\sqrt{3r_\text{ps}-4Q^2}}.$$ Due to the complexity of the system, we were unable to derive an analytical expression for the strong deflection coefficient $\bar{b}$; instead, we calculated it using the available numerical methods. The behaviour of the strong field deflection angle \eqref{sdef2} depicted in Fig. \ref{plot3}, demonstrates that the deflection angle for the charged \emph{LQGBHs} (solid curves) is greater than that for RN BHs (dashed curves) for $u>u_\text{ps}$ despite having the same divergence point at $u=u_\text{ps}$. The deviation in the deflection angle increases with increasing charge $Q$. We keep Barbero-Immirzi parameter $\gamma=\sqrt{3}/6$ for the results shown in Fig. \ref{plot3}.

Next, we examine the strong gravitational lensing observables obtained from the lens equation together with the photon deflection angle. The lens equation serves as a geometric framework that connects the observer $O$, the lens $L$, and the source $S$. Under the assumptions that the source and observer lie far from the lens, are nearly aligned, and are within a flat background spacetime \cite{Bozza:2002zj, Bozza:2008ev, Virbhadra:1999nm}, the lens equation can be expressed as
\begin{eqnarray}\label{lenseq}
\beta &=& \theta -\mathcal{D} \Delta\alpha_n,
\end{eqnarray} 
where $\Delta \alpha_{n}=\alpha(\theta)-2n\pi$ denotes the residual deflection angle after a photon loops around the lens by $2n\pi$. Here, $n$ is an integer ($n \in N$) with the condition $0 < \Delta \alpha_n \ll 1$. The quantities $\beta$ and $\theta$ indicate the angular positions of the source and image with respect to the optical axis. Furthermore, the projected distances from the observer and the source to the lens are given by $D_\text{OL}$ and $D_\text{LS}$, respectively, such that $D_\text{OS}=D_\text{OL}+D_\text{LS}$; while $\mathcal{D}$ is defined as their ratio, $D_\text{LS}/D_\text{OS}$. We now compute the angular position of the th relativistic image by substituting Eq.~(\ref{sdef2}) into the lens equation (\ref{lenseq}), which yields \cite{Bozza:2002zj}
\begin{equation}
    \theta_n=\theta^0_n+\frac{u_\text{ps} e_n(\beta-\theta^0_n)D_{OS}}{\bar{a}D_{LS}D_{OL}},
\end{equation} 
with
\begin{equation}
    e_n=\exp\left(\frac{\bar{b}}{\bar{a}}-\frac{2n\pi}{\bar{a}}\right),~~~~ \text{and}~~~~~\theta^0_n=\frac{u_\text{ps}(1+e_n)}{D_{OL}}
\end{equation}
where $\theta^0_n$ denotes the angular position of the image corresponding to a deflection angle of $2n\pi$. Under perfect alignment ($\beta=0$), the images form circular Einstein rings with radii:
\begin{equation}\label{einstein-rings}
\theta_n^E = \frac{u_{\text{ps}}}{D_{\text{OL}}}(1+e_n).
\end{equation}
The bending of light due to the lensing mass modifies the cross-sectional area of the light beam, producing a magnification of the image. According to Liouville’s theorem, gravitational deflection preserves surface brightness, implying that the magnification is entirely determined by the ratio of the solid angle subtended by the image to that of the unlensed source. Thus, the magnification factor for the relativistic image takes the form \cite{Virbhadra:1999nm, Bozza:2002zj}
\begin{equation}\label{mag}
    \mu_n=\left(\frac{\beta}{\theta}\frac{d\beta}{d\theta}\right)^{-1}\Bigg|_{\theta_n^0}=\frac{u_\text{ps} e_n(1+e_n)D_{OS}}{\bar{a}\beta D_{LS}D^2_{OL}}.
\end{equation}

\begin{figure}
	\begin{centering}
		\begin{tabular}{c c}
		  \includegraphics[width=0.5\textwidth]{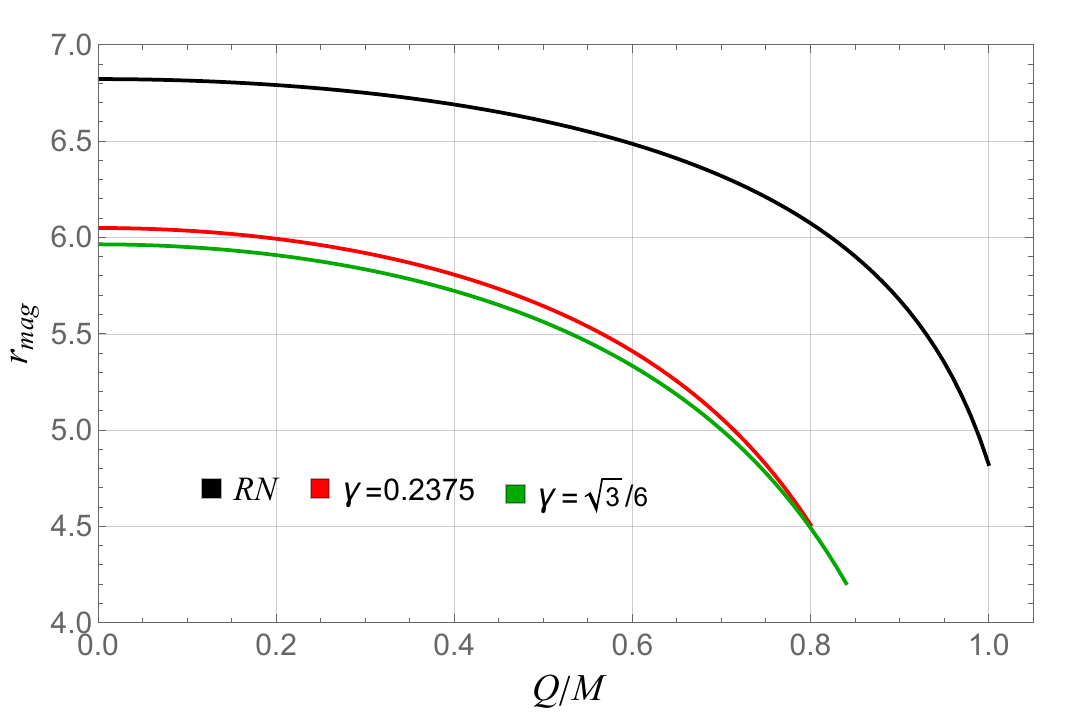}&
			\end{tabular}
	\end{centering}
	\caption{The strong--lensing observable $r_{\rm mag} = \mu_{1} / \sum_{n=2}^{\infty} \mu_{n}$ is plotted as a function of the charge $Q$ for RN and charged \emph{LQGBHs} with  $\gamma = 0.2375$ and $\gamma = \sqrt{3}/6$. 
LQG corrections reduce $r_{\rm mag}$, indicating that the first relativistic image becomes less dominant in brightness compared to higher-order images.}\label{plot5}		\end{figure}

\begin{table*}[htb!]
\resizebox{13cm}{!}{ 
 \begin{centering}	
	\begin{tabular}{|c|  c c| c c| c c| }
\hline
\multicolumn{1}{|c|}{}&
\multicolumn{2}{c|}{Sgr A*}&
\multicolumn{2}{c|}{M87*}&
\multicolumn{1}{c}{}&
\multicolumn{1}{c|}{}\\
 ~~~~{$Q$}& {$s_{LQG}$($\mu$as)}& ~~~{$s_{GR}$($\mu$as)} &{$s_{LQG}$($\mu$as)}& ~~~{$s_{GR}$($\mu$as)}&{$r_{LQG}$}& ~~~{$r_{GR}$} \\ 
\hline
~~~~$0$& $0.0759541$ &~~~$0.0329167$& $0.0571262$& ~~~$0.0247571$& $5.96397$ &~~~$6.82188$\\
 ~~~~$0.2$ & $0.0796531$ &~~~$0.0337314$& $0.0599082$& ~~~$0.0253699$ &$5.90807$ &~~~$6.79094$\\
 ~~~~$0.4$ & $0.0932424$ &~~~$0.036539$& $0.0701289$& ~~~$0.0274815$& $5.72239$ &~~~$6.68985$\\
 ~~~~$0.6$ & $0.129256$ &~~~$0.0429347$& $0.0972151$& ~~~$0.0322918$ & $5.33336$ &$ ~~~6.48575$\\
 ~~~~$0.8$ & $0.254267$ &~~~$0.0592217$& $0.191238$& ~~~$0.0445415$ & $4.49272$ &~~~$6.07378$\\
       \hline
	\end{tabular} 
\end{centering}}
	\caption{Lensing observables $s$ and $r_\text{mag}$ of SMBHs SgrA* and M87*. GR ($b_0\to1$) and LQG ($\gamma=\sqrt{3}/6$) predictions for $s$  and $r_\text{mag}$ are given for different values of charge $Q$.}\label{table1}
  \end{table*} 
  
  \begin{table*}[htb!]
\resizebox{\textwidth}{!}{ 
 \begin{centering}	
	\begin{tabular}{|c|  c| c c c c| c c c c| }
\hline
\multicolumn{2}{|c|}{Parameters}&
\multicolumn{4}{c|}{Sgr A*}&
\multicolumn{4}{c|}{M87*}\\
\hline
{$\mathcal{D}$ } & {$Q$}& {$\mu_{1p,LQG}$}& {$\mu_{2p,LQG}$} & {$\mu_{1p,GR}$ }& {$\mu_{2p,GR}$} & {$\mu_{1p,LQG}$}& {$\mu_{2p,LQG}$} & {$\mu_{1p,GR}$ }& {$\mu_{2p,GR}$} \\ \hline
\multirow{5}{*}{$0.01$}&$0$& $8.49175\times10^{-11}$ &$3.48467\times10^{-13}$& $4.20264\times10^{-11}$& $7.83839\times10^{-14}$& $4.80357\times10^{-11}$ &$1.97119\times10^{-13}$& $2.37733\times10^{-11}$& $4.43398\times10^{-14}$\\
& $0.2$ & $8.76396\times10^{-11}$ &$3.78578\times10^{-13}$& $4.25848\times10^{-11}$& $4.6226\times10^{-14}$ &$4.95756\times10^{-11}$ &$2.14152\times10^{-13}$& $2.40892\times10^{-11}$& $7.83839\times10^{-14}$\\
& $0.4$ & $9.73407\times10^{-11}$ &$4.98615\times10^{-13}$& $4.44955\times10^{-11}$& $9.37038\times10^{-14}$& $5.50633\times10^{-11}$ &$2.82054\times10^{-13}$& $2.517\times10^{-11}$& $5.30059\times10^{-14}$\\
& $0.6$ & $1.21122\times10^{-10}$ &$8.86377\times10^{-13}$& $4.87553\times10^{-11}$& $1.2387\times10^{-13}$ & $6.85158\times10^{-11}$ &$5.01402\times10^{-13}$& $2.75797\times10^{-11}$& $7.00704\times10^{-14}$\\
& $0.8$ & $1.88876\times10^{-10}$ &$1.31921\times10^{-12}$& $5.8974\times10^{-11}$& $2.18795\times10^{-13}$ 
& $1.06842\times10^{-10}$ &$7.46242\times10^{-13}$& $3.33601\times10^{-11}$& $1.23767\times10^{-13}$\\
\hline
\multirow{5}{*}{$0.1$}&$0$& $8.49175\times10^{-12}$ &$3.48467\times10^{-14}$& $4.20264\times10^{-12}$& $7.83839\times10^{-15}$& $4.80357\times10^{-12}$& $1.97119\times10^{-14}$&$2.37733\times10^{-12}$&$4.43398\times10^{-15}$\\
& $0.2$ &$8.76396\times10^{-12}$ &$3.78578\times10^{-14}$& $4.25848\times10^{-12}$& $8.17183\times10^{-15}$ &$4.95756\times10^{-12}$& $2.14152\times10^{-14}$& $2.40892\times10^{-12}$&$4.6226\times10^{-15}$\\
& $0.4$ & $9.73407\times10^{-12}$ &$4.98615\times10^{-14}$& $4.44955\times10^{-12}$& $ 9.37038\times10^{-15}$ & $5.50633\times10^{-12}$& $2.82054\times10^{-14}$ &$2.517\times10^{-12}$&$5.30059\times10^{-15}$\\
& $0.6$ & $1.21122\times10^{-11}$ &$8.86377\times10^{-14}$& $4.87553\times10^{-12}$& $1.2387\times10^{-14}$ & $6.85158\times10^{-12}$& $5.01402\times10^{-14}$ &$2.75797\times10^{-12}$&$7.00704\times10^{-15}$\\
& $0.8$ & $1.88876\times10^{-11}$ &$1.31921\times10^{-13}$& $5.8974\times10^{-12}$& $2.18795\times10^{-14}$ & $1.06842\times10^{-11}$& $7.46242\times10^{-14}$ &$3.33601\times10^{-12}$&$1.23767\times10^{-14}$\\
\hline
\multirow{5}{*}{$0.5$}&$0$& $1.69835\times10^{-12}$ &$6.96934\times10^{-15}$& $8.40527\times10^{-13}$& $1.56768\times10^{-15}$ & $9.60714\times10^{-13}$& $3.94238\times10^{-15}$ &$4.75466\times10^{-13}$ &$8.86797\times10^{-16}$\\
& $0.2$ &$1.75279\times10^{-12}$ &$7.57156\times10^{-15}$& $8.51696\times10^{-13}$& $1.63437\times10^{-15}$ & $9.91511\times10^{-13}$& $4.28304\times10^{-15}$ &$4.81784\times10^{-13}$ &$9.2452\times10^{-16}$\\
& $0.4$ & $1.94681\times10^{-12}$ &$9.9723\times10^{-15}$& $8.89909\times10^{-13}$& $ 1.87408\times10^{-15}$ & $1.10127\times10^{-12}$& $5.64109\times10^{-15}$ &$5.034\times10^{-13}$ &$1.06012\times10^{-15}$\\
& $0.6$ & $2.42244\times10^{-12}$ &$1.77275\times10^{-14}$& $9.75106\times10^{-13}$& $2.47741\times10^{-15}$ & $1.37032\times10^{-12}$& $1.0028\times10^{-14}$ &$5.51594\times10^{-13}$ &$1.40141\times10^{-15}$\\
& $0.8$ & $3.77751\times10^{-12}$ &$2.63841\times10^{-14}$& $1.17948\times10^{-12}$& $4.3759\times10^{-15}$ & $2.13685\times10^{-12}$& $1.49248\times10^{-14}$ &$6.67202\times10^{-13}$ &$2.47534\times10^{-15}$\\
       \hline
	\end{tabular} 
\end{centering}}
	\caption{Estimates for the magnification $\mu_n$ of the first two relativistic images of SMBHs SgrA* and M87* for different values of $\mathcal{D}=D_\text{LS}/D_\text{OS}$. GR ($b_0\to1$) and LQG ($\gamma=\sqrt{3}/6$) predictions for magnifications $\mu_n$ are given for different values of charge $Q$.}\label{table2}
\end{table*}

\begin{figure*}
	\begin{centering}
		\begin{tabular}{c c}
		  \includegraphics[width=0.5\textwidth]{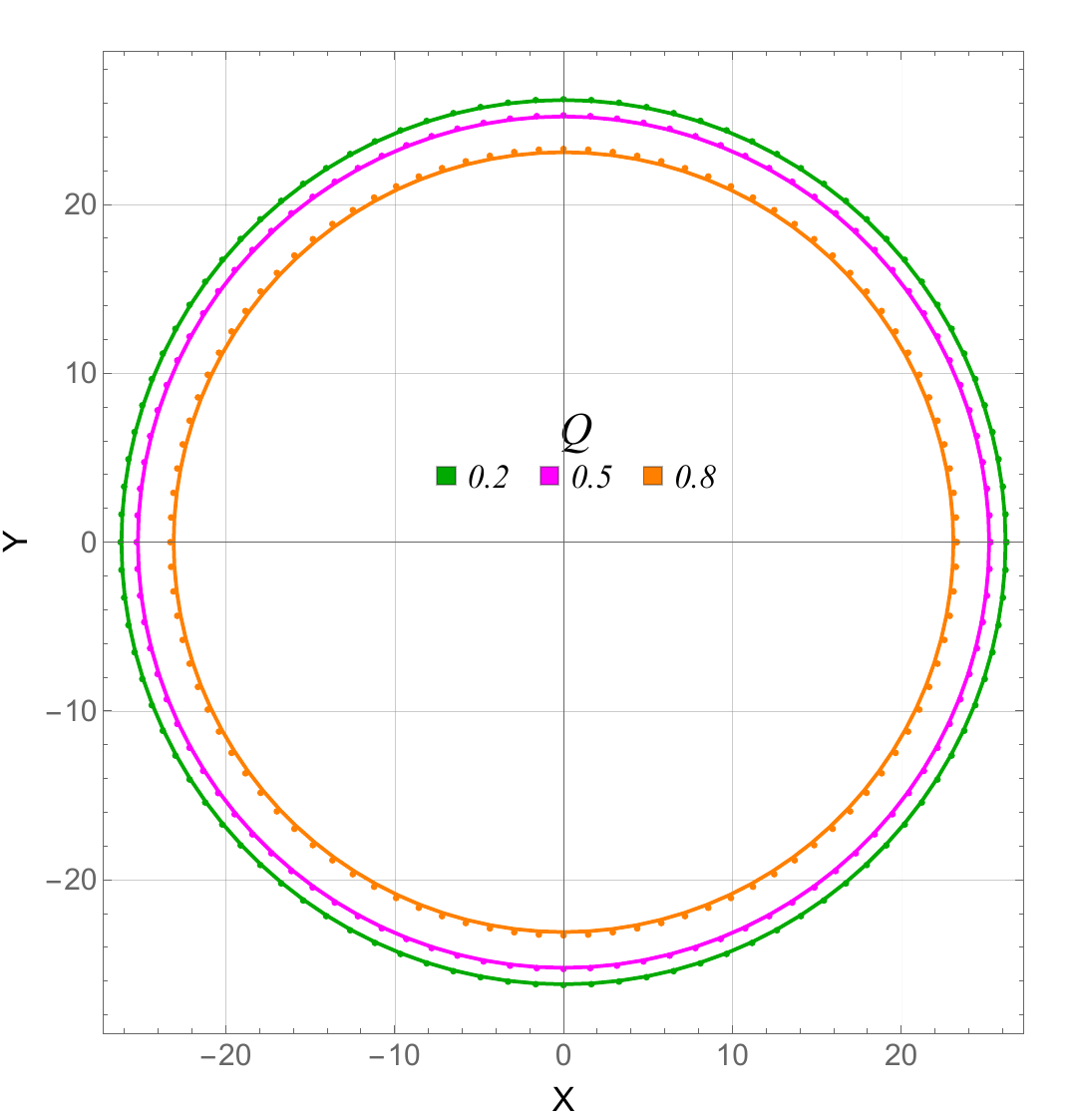}&
			\includegraphics[width=0.5\textwidth]{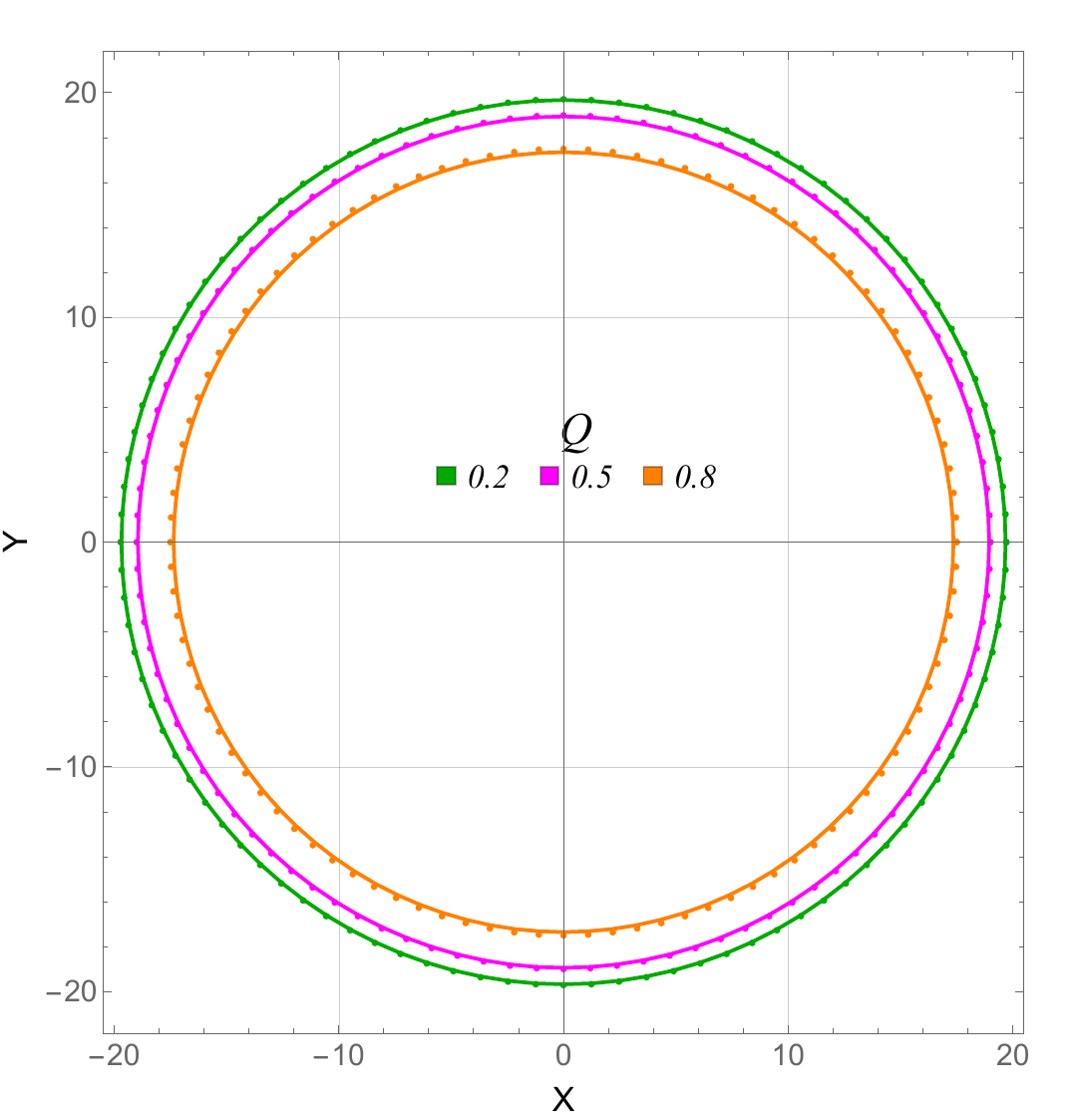}
			\end{tabular}
	\end{centering}
	\caption{Einstein rings in the strong-field regime for the supermassive black holes Sgr A$^*$ (left) and M87$^*$ (right), modeled as charged \emph{LQGBH}  (solid rings) and RN black holes (dotted rings), for different values of the charge $Q$. The Barbero--Immirzi parameter is fixed at $\gamma = \sqrt{3}/6$.}
 \label{plot6}		
\end{figure*}

For the special case in which the source lies exactly on the optical axis ($\beta\to0$), Eq.~(\ref{mag}) diverges, indicating that the likelihood of observing a gravitational lensing image is maximized under perfect alignment. Moreover, the magnification decreases exponentially with the image order $n$, implying that the first relativistic image $\theta_1$ is the brightest. The angular isolation of the outermost image ($n=1$) from the densely clustered sequence of images ($n=2,3,....,\infty$) allows one to define the following observable quantities \cite{Bozza:2002zj}:
\begin{eqnarray}\label{observable}
    && \theta_{\infty}=\frac{u_\text{ps}}{D_{OL}},
    ~~~~~ s=\theta_1-\theta_{\infty}\approx \theta_{\infty} \exp\left[\frac{\bar{b}-2\pi}{\bar{a}}\right],
    \nonumber\\&& r_\text{mag}=\frac{\mu_1}{\sum^{\infty}_{n=2} \mu_{n}}\approx\frac{5\pi}{\bar{a}\log(10)}. 
\end{eqnarray}
In this framework, $\theta_{\infty}$ represents the limiting angular position where the set of higher-order images accumulates, $s$ describes the observable angular separation between the first relativistic image and the cluster of subsequent images, and $r_\text{mag}$ quantifies the relative flux ratio between the first observable image and all of the remaining packed images.

\subsection{Lensing observables analysis of supermassive Sgr A* and M87* BHs}\label{3C}

The EHT targets, supermassive BHs M87* and Sgr A*, are modeled as charged BHs in LQG to estimate their lensing observables. These results are then compared with those for standard RN BHs. According to the latest astronomical observations, the estimated mass of M87* is $\left(6.5\pm0.7\right)\times{10}^9M_\odot$, with a distance of $d=16.8$ MPc \cite{EventHorizonTelescope:2019ggy}. The estimated mass of Sgr A* is $4.28_{\pm0.10}^{\pm0.21}\times{10}^6M_\odot$ \cite{2017ApJ...837...30G}, and its distance is $d=8.32^{\pm0.07}_{\pm0.14}$ KPc \cite{2017ApJ...837...30G}. 
  
The relativistic angular image position $\theta_{\infty}$ is independent of the Barbero-Immirzi parameter $\gamma$ and the polymerisation parameter $\delta_b$, and hence $\theta_{\infty}$ for charged \emph{LQGBHs} and RN BHs has the same value, which depends only on $Q$. However, the extremal value of $Q$ for both cases is not the same. When we analyse the behaviour of the angular separation observable $s$, we find that the $s$ for charged \emph{LQGBHs} is greater than that for RN BHs. It increases with $\gamma$ (cf. Fig. \ref{plot4}), while the relative flux ratio $r_{mag}$, which is depicted in Fig. \ref{plot5}, exhibits opposite behaviour to that of $s$, e.i. it has a lower value for charged \emph{LQGBHs} than that for RN BHs, and it decreases with $\gamma$. The numerical results of $s$ and $r_{mag}$ for charged \emph{LQGBHs} ($\gamma=\sqrt{3}/6$) and RN BHs have been given in Table \ref{table1}, where subscripts LQG and GR represent charged \emph{LQGBHs} and RN BHs, respectively. Table \ref{table2} shows the magnifications of $n$-th order images of charged \emph{LQGBHs} and compares them with those of the RN BHs for Sgr A* and M87*. We found that higher-order images are much less bright, and their magnification decreases as the distance ratio $\mathcal{D}$ increases. The comparative analysis image magnification reveals that charged \emph{LQGBHs} make brighter images than those of RN BHs (cf. Table \ref{table2}). By using Eq. \eqref{einstein-rings}, we plotted Einstein's rings in strong field limits for SMBHs modelled as charged \emph{LQGBHs} and RN BHs for different values of $Q$ in Fig. \ref{plot6}. The LQG corrections slightly enlarge the radius of Einstein's ring and the difference between the Einstein's rings of charged \emph{LQGBHs} and RN BHs becomes more prominent as we increase $Q$ (cf. Fig. \ref{plot6}).
 
\section{CONCLUSION}\label{sec4}
 
The search to unify GR with quantum gravity remains one of the most profound challenges in theoretical physics. While black holes represent a key testing ground for such a unified theory, direct observational signatures of quantum gravity have remained elusive. The advent of gravitational-wave and horizon-scale imaging by the LIGO-Virgo-KAGRA collaborations, as well as by the EHT, has now opened a new laboratory into the strong-gravity regime. In this context, strong gravitational lensing serves as a powerful probe of the underlying spacetime geometry. This work is motivated by the potential to use these precise lensing observables to distinguish between GR and LQG BHs, thereby turning astrophysical BHs into laboratories.

In particular, we investigate strong gravitational lensing by charged \emph{LQGBHs} and compare the results with those of classical RN BHs. In the strong-field regime, our analysis revealed several unique characteristics brought about by LQG corrections that may serve as observational indicators of quantum gravity.  The photon sphere radius $ r_{\text{ps}} $ and critical impact parameter $u_{\text{ps}} $ for charged \emph{LQGBHs} are identical to those of RN BHs. However, the allowed range of charge $Q$ is significantly restricted due to LQG constraints. For instance, with $\gamma = \sqrt{3}/6 $ and $ M = 1 $, the maximum allowable charge is $Q_{\text{max}} = 0.840896 $, which is lower than the classical GR limit of $ Q = 1 $.

The weak-field deflection angle for \emph{LQGBHs} is enhanced compared to RN, and this enhancement increases with charge $Q$. For example, at  $Q = 0.8$ and $ \gamma = \sqrt{3}/6$, the deviation $\delta\alpha_D(u) $ is positive and increases with decreasing impact parameter $ u $. The angular radius of the Einstein ring $ \theta_E $ for  \emph{LQGBHs} increases with both $Q$ and $\gamma $, in contrast to RN BHs, where $ \theta_E $ decreases with $Q$. For the galaxy ESO325-G004, the LQG model indicates $\theta_E $ should lie within the $1\sigma $ observational uncertainty of $2.85^{+0.55}_{-0.25}$ arcsec. 

The strong deflection angle coefficient $\bar{a} $ is derived analytically, while $\bar{b} $ is computed numerically. The deflection angle in the strong-field regime is larger for charged \emph{LQGBHs} than for RN BHs, with the deviation increasing with $Q$. For example, at $Q = 0.8$, the deflection angle $\alpha_D(u)$ is significantly higher in the LQG case. The angular separation $ s $ between the first relativistic image and the image cluster is larger for charged \emph{LQGBHs} than for RN BHs. For Sgr A* at $Q = 0.8$, $ s_{\text{LQG}} = 0.254267 \mu\text{as}$, compared to $s_{\text{GR}} = 0.0592217  \mu\text{as} $. The flux ratio $r_{\text{mag}} $ is lower for charged \emph{LQGBHs}. For Sgr A* at $Q = 0.8$, $r_{\text{LQG}} = 4.49272$, while $r_{\text{GR}} = 6.07378 $. The magnification $\mu_n$ is higher for charged \emph{LQGBHs}, indicating that they produce brighter images than their GR counterparts. For example, for Sgr A* with $ \mathcal{D} = 0.01 $ and $ Q = 0.8 $, the first-image magnification is $\mu_{1,\text{LQG}} = 1.88876 \times 10^{-10}$, compared to $\mu_{1,\text{GR}} = 5.8974 \times 10^{-11}$.

These findings demonstrate the potential of strong gravitational lensing as an effective method for differentiating between the geometries of classical and quantum-corrected black holes. Future ngEHT observation could place stringent constraints on LQG  Barbero–Immirzi parameter $ \gamma $ and the polymerization parameter $ \delta_b$, thereby providing unique insights into the quantum nature of gravity in the strong-field regime. This work opens several key avenues for future study. Extending the analysis to \textbf{rotating  \emph{LQGBHs}} is essential for realistic comparison with astrophysical observations from the EHT. Exploring the \textbf{multi-messenger link} between lensing observables and gravitational-wave quasinormal modes could furnish a complementary probe. Finally, investigating the lensing properties of the predicted \textbf{Planck-scale remnants} may offer insights into their potential role as dark matter constituents.

\section*{Acknowledgements}

SGG would like to thank the Inter-University Centre for Astronomy and Astrophysics (IUCAA), Pune, for its hospitality during the associateship visit, during which part of this work was conducted. This work is supported by the Zhejiang Provincial Natural Science Foundation of China under Grants No.~LR21A050001 and No.~LY20A050002, the National Natural Science Foundation of China under Grants No.~12275238, No. W2433018, and No. 11675143, the National Key Research and Development Program of China under Grant No. 2020YFC2201503, and the Fundamental Research Funds for the Provincial Universities of Zhejiang in China under Grant No.~RF-A2019015.

\bibliographystyle{apsrev4-1}
\bibliography{LQG.bib}

\begin{thebibliography}{114}%
\makeatletter
\providecommand \@ifxundefined [1]{%
 \@ifx{#1\undefined}
}%
\providecommand \@ifnum [1]{%
 \ifnum #1\expandafter \@firstoftwo
 \else \expandafter \@secondoftwo
 \fi
}%
\providecommand \@ifx [1]{%
 \ifx #1\expandafter \@firstoftwo
 \else \expandafter \@secondoftwo
 \fi
}%
\providecommand \natexlab [1]{#1}%
\providecommand \enquote  [1]{``#1''}%
\providecommand \bibnamefont  [1]{#1}%
\providecommand \bibfnamefont [1]{#1}%
\providecommand \citenamefont [1]{#1}%
\providecommand \href@noop [0]{\@secondoftwo}%
\providecommand \href [0]{\begingroup \@sanitize@url \@href}%
\providecommand \@href[1]{\@@startlink{#1}\@@href}%
\providecommand \@@href[1]{\endgroup#1\@@endlink}%
\providecommand \@sanitize@url [0]{\catcode `\\12\catcode `\$12\catcode `\&12\catcode `\#12\catcode `\^12\catcode `\_12\catcode `\%12\relax}%
\providecommand \@@startlink[1]{}%
\providecommand \@@endlink[0]{}%
\providecommand \url  [0]{\begingroup\@sanitize@url \@url }%
\providecommand \@url [1]{\endgroup\@href {#1}{\urlprefix }}%
\providecommand \urlprefix  [0]{URL }%
\providecommand \Eprint [0]{\href }%
\providecommand \doibase [0]{http://dx.doi.org/}%
\providecommand \selectlanguage [0]{\@gobble}%
\providecommand \bibinfo  [0]{\@secondoftwo}%
\providecommand \bibfield  [0]{\@secondoftwo}%
\providecommand \translation [1]{[#1]}%
\providecommand \BibitemOpen [0]{}%
\providecommand \bibitemStop [0]{}%
\providecommand \bibitemNoStop [0]{.\EOS\space}%
\providecommand \EOS [0]{\spacefactor3000\relax}%
\providecommand \BibitemShut  [1]{\csname bibitem#1\endcsname}%
\let\auto@bib@innerbib\@empty
\bibitem [{\citenamefont {Borges}\ \emph {et~al.}(2024{\natexlab{a}})\citenamefont {Borges}, \citenamefont {Baranov}, \citenamefont {Sobrinho},\ and\ \citenamefont {Carneiro}}]{Borges:2023fog}%
  \BibitemOpen
  \bibfield  {author} {\bibinfo {author} {\bibfnamefont {H.~A.}\ \bibnamefont {Borges}}, \bibinfo {author} {\bibfnamefont {I.~P.~R.}\ \bibnamefont {Baranov}}, \bibinfo {author} {\bibfnamefont {F.~C.}\ \bibnamefont {Sobrinho}}, \ and\ \bibinfo {author} {\bibfnamefont {S.}~\bibnamefont {Carneiro}},\ }\href {\doibase 10.1140/epjc/s10052-024-13235-1} {\bibfield  {journal} {\bibinfo  {journal} {Eur. Phys. J. C}\ }\textbf {\bibinfo {volume} {84}},\ \bibinfo {pages} {63} (\bibinfo {year} {2024}{\natexlab{a}})},\ \Eprint {http://arxiv.org/abs/2310.01560} {arXiv:2310.01560 [gr-qc]} \BibitemShut {NoStop}%
\bibitem [{\citenamefont {Abbott}\ \emph {et~al.}(2016)\citenamefont {Abbott} \emph {et~al.}}]{LIGOScientific:2016aoc}%
  \BibitemOpen
  \bibfield  {author} {\bibinfo {author} {\bibfnamefont {B.~P.}\ \bibnamefont {Abbott}} \emph {et~al.} (\bibinfo {collaboration} {LIGO Scientific, Virgo}),\ }\href {\doibase 10.1103/PhysRevLett.116.061102} {\bibfield  {journal} {\bibinfo  {journal} {Phys. Rev. Lett.}\ }\textbf {\bibinfo {volume} {116}},\ \bibinfo {pages} {061102} (\bibinfo {year} {2016})},\ \Eprint {http://arxiv.org/abs/1602.03837} {arXiv:1602.03837 [gr-qc]} \BibitemShut {NoStop}%
\bibitem [{\citenamefont {Abbott}\ \emph {et~al.}(2021)\citenamefont {Abbott} \emph {et~al.}}]{LIGOScientific:2021djp}%
  \BibitemOpen
  \bibfield  {author} {\bibinfo {author} {\bibfnamefont {R.}~\bibnamefont {Abbott}} \emph {et~al.} (\bibinfo {collaboration} {LIGO Scientific, VIRGO, KAGRA}),\ }\href@noop {} {\  (\bibinfo {year} {2021})},\ \Eprint {http://arxiv.org/abs/2111.03606} {arXiv:2111.03606 [gr-qc]} \BibitemShut {NoStop}%
\bibitem [{\citenamefont {Akiyama}\ \emph {et~al.}(2019{\natexlab{a}})\citenamefont {Akiyama} \emph {et~al.}}]{EventHorizonTelescope:2019dse}%
  \BibitemOpen
  \bibfield  {author} {\bibinfo {author} {\bibfnamefont {K.}~\bibnamefont {Akiyama}} \emph {et~al.} (\bibinfo {collaboration} {Event Horizon Telescope}),\ }\href {\doibase 10.3847/2041-8213/ab0ec7} {\bibfield  {journal} {\bibinfo  {journal} {Astrophys. J. Lett.}\ }\textbf {\bibinfo {volume} {875}},\ \bibinfo {pages} {L1} (\bibinfo {year} {2019}{\natexlab{a}})},\ \Eprint {http://arxiv.org/abs/1906.11238} {arXiv:1906.11238 [astro-ph.GA]} \BibitemShut {NoStop}%
\bibitem [{\citenamefont {Akiyama}\ \emph {et~al.}(2022)\citenamefont {Akiyama} \emph {et~al.}}]{EventHorizonTelescope:2022wkp}%
  \BibitemOpen
  \bibfield  {author} {\bibinfo {author} {\bibfnamefont {K.}~\bibnamefont {Akiyama}} \emph {et~al.} (\bibinfo {collaboration} {Event Horizon Telescope}),\ }\href {\doibase 10.3847/2041-8213/ac6674} {\bibfield  {journal} {\bibinfo  {journal} {Astrophys. J. Lett.}\ }\textbf {\bibinfo {volume} {930}},\ \bibinfo {pages} {L12} (\bibinfo {year} {2022})},\ \Eprint {http://arxiv.org/abs/2311.08680} {arXiv:2311.08680 [astro-ph.HE]} \BibitemShut {NoStop}%
\bibitem [{\citenamefont {Bartelmann}(2010)}]{Bartelmann:2010fz}%
  \BibitemOpen
  \bibfield  {author} {\bibinfo {author} {\bibfnamefont {M.}~\bibnamefont {Bartelmann}},\ }\href {\doibase 10.1088/0264-9381/27/23/233001} {\bibfield  {journal} {\bibinfo  {journal} {Class. Quant. Grav.}\ }\textbf {\bibinfo {volume} {27}},\ \bibinfo {pages} {233001} (\bibinfo {year} {2010})},\ \Eprint {http://arxiv.org/abs/1010.3829} {arXiv:1010.3829 [astro-ph.CO]} \BibitemShut {NoStop}%
\bibitem [{\citenamefont {Weinberg}(1972)}]{Weinberg:1972kfs}%
  \BibitemOpen
  \bibfield  {author} {\bibinfo {author} {\bibfnamefont {S.}~\bibnamefont {Weinberg}},\ }\href@noop {} {\emph {\bibinfo {title} {{Gravitation and Cosmology}: {Principles and Applications of the General Theory of Relativity}}}}\ (\bibinfo  {publisher} {John Wiley and Sons},\ \bibinfo {address} {New York},\ \bibinfo {year} {1972})\BibitemShut {NoStop}%
\bibitem [{\citenamefont {Schneider}\ \emph {et~al.}(1992)\citenamefont {Schneider}, \citenamefont {Ehlers},\ and\ \citenamefont {Falco}}]{Schneider:1992bmb}%
  \BibitemOpen
  \bibfield  {author} {\bibinfo {author} {\bibfnamefont {P.}~\bibnamefont {Schneider}}, \bibinfo {author} {\bibfnamefont {J.}~\bibnamefont {Ehlers}}, \ and\ \bibinfo {author} {\bibfnamefont {E.~E.}\ \bibnamefont {Falco}},\ }\href {\doibase 10.1007/978-3-662-03758-4} {\emph {\bibinfo {title} {{Gravitational Lenses}}}},\ Astronomy and Astrophysics Library\ (\bibinfo  {publisher} {Springer},\ \bibinfo {year} {1992})\BibitemShut {NoStop}%
\bibitem [{\citenamefont {Virbhadra}\ and\ \citenamefont {Ellis}(2000)}]{Virbhadra:1999nm}%
  \BibitemOpen
  \bibfield  {author} {\bibinfo {author} {\bibfnamefont {K.~S.}\ \bibnamefont {Virbhadra}}\ and\ \bibinfo {author} {\bibfnamefont {G.~F.~R.}\ \bibnamefont {Ellis}},\ }\href {\doibase 10.1103/PhysRevD.62.084003} {\bibfield  {journal} {\bibinfo  {journal} {Phys. Rev. D}\ }\textbf {\bibinfo {volume} {62}},\ \bibinfo {pages} {084003} (\bibinfo {year} {2000})},\ \Eprint {http://arxiv.org/abs/astro-ph/9904193} {arXiv:astro-ph/9904193} \BibitemShut {NoStop}%
\bibitem [{\citenamefont {Virbhadra}(2009)}]{Virbhadra:2008ws}%
  \BibitemOpen
  \bibfield  {author} {\bibinfo {author} {\bibfnamefont {K.~S.}\ \bibnamefont {Virbhadra}},\ }\href {\doibase 10.1103/PhysRevD.79.083004} {\bibfield  {journal} {\bibinfo  {journal} {Phys. Rev. D}\ }\textbf {\bibinfo {volume} {79}},\ \bibinfo {pages} {083004} (\bibinfo {year} {2009})},\ \Eprint {http://arxiv.org/abs/0810.2109} {arXiv:0810.2109 [gr-qc]} \BibitemShut {NoStop}%
\bibitem [{\citenamefont {Claudel}\ \emph {et~al.}(2001)\citenamefont {Claudel}, \citenamefont {Virbhadra},\ and\ \citenamefont {Ellis}}]{Claudel:2000yi}%
  \BibitemOpen
  \bibfield  {author} {\bibinfo {author} {\bibfnamefont {C.-M.}\ \bibnamefont {Claudel}}, \bibinfo {author} {\bibfnamefont {K.~S.}\ \bibnamefont {Virbhadra}}, \ and\ \bibinfo {author} {\bibfnamefont {G.~F.~R.}\ \bibnamefont {Ellis}},\ }\href {\doibase 10.1063/1.1308507} {\bibfield  {journal} {\bibinfo  {journal} {J. Math. Phys.}\ }\textbf {\bibinfo {volume} {42}},\ \bibinfo {pages} {818} (\bibinfo {year} {2001})},\ \Eprint {http://arxiv.org/abs/gr-qc/0005050} {arXiv:gr-qc/0005050} \BibitemShut {NoStop}%
\bibitem [{\citenamefont {{Darwin}}(1959)}]{1959RSPSA.249..180D}%
  \BibitemOpen
  \bibfield  {author} {\bibinfo {author} {\bibfnamefont {C.}~\bibnamefont {{Darwin}}},\ }\href {\doibase 10.1098/rspa.1959.0015} {\bibfield  {journal} {\bibinfo  {journal} {Proceedings of the Royal Society of London Series A}\ }\textbf {\bibinfo {volume} {249}},\ \bibinfo {pages} {180} (\bibinfo {year} {1959})}\BibitemShut {NoStop}%
\bibitem [{\citenamefont {Bozza}(2002)}]{Bozza:2002zj}%
  \BibitemOpen
  \bibfield  {author} {\bibinfo {author} {\bibfnamefont {V.}~\bibnamefont {Bozza}},\ }\href {\doibase 10.1103/PhysRevD.66.103001} {\bibfield  {journal} {\bibinfo  {journal} {Phys. Rev. D}\ }\textbf {\bibinfo {volume} {66}},\ \bibinfo {pages} {103001} (\bibinfo {year} {2002})},\ \Eprint {http://arxiv.org/abs/gr-qc/0208075} {arXiv:gr-qc/0208075} \BibitemShut {NoStop}%
\bibitem [{\citenamefont {Bozza}\ \emph {et~al.}(2001)\citenamefont {Bozza}, \citenamefont {Capozziello}, \citenamefont {Iovane},\ and\ \citenamefont {Scarpetta}}]{Bozza:2001xd}%
  \BibitemOpen
  \bibfield  {author} {\bibinfo {author} {\bibfnamefont {V.}~\bibnamefont {Bozza}}, \bibinfo {author} {\bibfnamefont {S.}~\bibnamefont {Capozziello}}, \bibinfo {author} {\bibfnamefont {G.}~\bibnamefont {Iovane}}, \ and\ \bibinfo {author} {\bibfnamefont {G.}~\bibnamefont {Scarpetta}},\ }\href {\doibase 10.1023/A:1012292927358} {\bibfield  {journal} {\bibinfo  {journal} {Gen. Rel. Grav.}\ }\textbf {\bibinfo {volume} {33}},\ \bibinfo {pages} {1535} (\bibinfo {year} {2001})},\ \Eprint {http://arxiv.org/abs/gr-qc/0102068} {arXiv:gr-qc/0102068} \BibitemShut {NoStop}%
\bibitem [{\citenamefont {Bozza}\ and\ \citenamefont {Mancini}(2004)}]{Bozza:2003cp}%
  \BibitemOpen
  \bibfield  {author} {\bibinfo {author} {\bibfnamefont {V.}~\bibnamefont {Bozza}}\ and\ \bibinfo {author} {\bibfnamefont {L.}~\bibnamefont {Mancini}},\ }\href {\doibase 10.1023/B:GERG.0000010486.58026.4f} {\bibfield  {journal} {\bibinfo  {journal} {Gen. Rel. Grav.}\ }\textbf {\bibinfo {volume} {36}},\ \bibinfo {pages} {435} (\bibinfo {year} {2004})},\ \Eprint {http://arxiv.org/abs/gr-qc/0305007} {arXiv:gr-qc/0305007} \BibitemShut {NoStop}%
\bibitem [{\citenamefont {Seto}(2004)}]{Seto:2003iw}%
  \BibitemOpen
  \bibfield  {author} {\bibinfo {author} {\bibfnamefont {N.}~\bibnamefont {Seto}},\ }\href {\doibase 10.1103/PhysRevD.69.022002} {\bibfield  {journal} {\bibinfo  {journal} {Phys. Rev. D}\ }\textbf {\bibinfo {volume} {69}},\ \bibinfo {pages} {022002} (\bibinfo {year} {2004})},\ \Eprint {http://arxiv.org/abs/astro-ph/0305605} {arXiv:astro-ph/0305605} \BibitemShut {NoStop}%
\bibitem [{\citenamefont {Vazquez}\ and\ \citenamefont {Esteban}(2004)}]{Vazquez:2003zm}%
  \BibitemOpen
  \bibfield  {author} {\bibinfo {author} {\bibfnamefont {S.~E.}\ \bibnamefont {Vazquez}}\ and\ \bibinfo {author} {\bibfnamefont {E.~P.}\ \bibnamefont {Esteban}},\ }\href {\doibase 10.1393/ncb/i2004-10121-y} {\bibfield  {journal} {\bibinfo  {journal} {Nuovo Cim. B}\ }\textbf {\bibinfo {volume} {119}},\ \bibinfo {pages} {489} (\bibinfo {year} {2004})},\ \Eprint {http://arxiv.org/abs/gr-qc/0308023} {arXiv:gr-qc/0308023} \BibitemShut {NoStop}%
\bibitem [{\citenamefont {Bozza}\ \emph {et~al.}(2005)\citenamefont {Bozza}, \citenamefont {De~Luca}, \citenamefont {Scarpetta},\ and\ \citenamefont {Sereno}}]{Bozza:2005tg}%
  \BibitemOpen
  \bibfield  {author} {\bibinfo {author} {\bibfnamefont {V.}~\bibnamefont {Bozza}}, \bibinfo {author} {\bibfnamefont {F.}~\bibnamefont {De~Luca}}, \bibinfo {author} {\bibfnamefont {G.}~\bibnamefont {Scarpetta}}, \ and\ \bibinfo {author} {\bibfnamefont {M.}~\bibnamefont {Sereno}},\ }\href {\doibase 10.1103/PhysRevD.72.083003} {\bibfield  {journal} {\bibinfo  {journal} {Phys. Rev. D}\ }\textbf {\bibinfo {volume} {72}},\ \bibinfo {pages} {083003} (\bibinfo {year} {2005})},\ \Eprint {http://arxiv.org/abs/gr-qc/0507137} {arXiv:gr-qc/0507137} \BibitemShut {NoStop}%
\bibitem [{\citenamefont {Bozza}\ \emph {et~al.}(2006)\citenamefont {Bozza}, \citenamefont {De~Luca},\ and\ \citenamefont {Scarpetta}}]{Bozza:2006nm}%
  \BibitemOpen
  \bibfield  {author} {\bibinfo {author} {\bibfnamefont {V.}~\bibnamefont {Bozza}}, \bibinfo {author} {\bibfnamefont {F.}~\bibnamefont {De~Luca}}, \ and\ \bibinfo {author} {\bibfnamefont {G.}~\bibnamefont {Scarpetta}},\ }\href {\doibase 10.1103/PhysRevD.74.063001} {\bibfield  {journal} {\bibinfo  {journal} {Phys. Rev. D}\ }\textbf {\bibinfo {volume} {74}},\ \bibinfo {pages} {063001} (\bibinfo {year} {2006})},\ \Eprint {http://arxiv.org/abs/gr-qc/0604093} {arXiv:gr-qc/0604093} \BibitemShut {NoStop}%
\bibitem [{\citenamefont {Gyulchev}\ and\ \citenamefont {Yazadjiev}(2007{\natexlab{a}})}]{Gyulchev:2006zg}%
  \BibitemOpen
  \bibfield  {author} {\bibinfo {author} {\bibfnamefont {G.~N.}\ \bibnamefont {Gyulchev}}\ and\ \bibinfo {author} {\bibfnamefont {S.~S.}\ \bibnamefont {Yazadjiev}},\ }\href {\doibase 10.1103/PhysRevD.75.023006} {\bibfield  {journal} {\bibinfo  {journal} {Phys. Rev. D}\ }\textbf {\bibinfo {volume} {75}},\ \bibinfo {pages} {023006} (\bibinfo {year} {2007}{\natexlab{a}})},\ \Eprint {http://arxiv.org/abs/gr-qc/0611110} {arXiv:gr-qc/0611110} \BibitemShut {NoStop}%
\bibitem [{\citenamefont {Gyulchev}\ and\ \citenamefont {Yazadjiev}(2007{\natexlab{b}})}]{Gyulchev:2007zz}%
  \BibitemOpen
  \bibfield  {author} {\bibinfo {author} {\bibfnamefont {G.~N.}\ \bibnamefont {Gyulchev}}\ and\ \bibinfo {author} {\bibfnamefont {S.~S.}\ \bibnamefont {Yazadjiev}},\ }\href {\doibase 10.1063/1.2733078} {\bibfield  {journal} {\bibinfo  {journal} {AIP Conf. Proc.}\ }\textbf {\bibinfo {volume} {899}},\ \bibinfo {pages} {145} (\bibinfo {year} {2007}{\natexlab{b}})}\BibitemShut {NoStop}%
\bibitem [{\citenamefont {Virbhadra}\ and\ \citenamefont {Ellis}(2002)}]{Virbhadra:2002ju}%
  \BibitemOpen
  \bibfield  {author} {\bibinfo {author} {\bibfnamefont {K.~S.}\ \bibnamefont {Virbhadra}}\ and\ \bibinfo {author} {\bibfnamefont {G.~F.~R.}\ \bibnamefont {Ellis}},\ }\href {\doibase 10.1103/PhysRevD.65.103004} {\bibfield  {journal} {\bibinfo  {journal} {Phys. Rev. D}\ }\textbf {\bibinfo {volume} {65}},\ \bibinfo {pages} {103004} (\bibinfo {year} {2002})}\BibitemShut {NoStop}%
\bibitem [{\citenamefont {Atamurotov}\ and\ \citenamefont {Ghosh}(2022)}]{Atamurotov:2022srw}%
  \BibitemOpen
  \bibfield  {author} {\bibinfo {author} {\bibfnamefont {F.}~\bibnamefont {Atamurotov}}\ and\ \bibinfo {author} {\bibfnamefont {S.~G.}\ \bibnamefont {Ghosh}},\ }\href {\doibase 10.1140/epjp/s13360-022-02885-3} {\bibfield  {journal} {\bibinfo  {journal} {Eur. Phys. J. Plus}\ }\textbf {\bibinfo {volume} {137}},\ \bibinfo {pages} {662} (\bibinfo {year} {2022})}\BibitemShut {NoStop}%
\bibitem [{\citenamefont {Younas}\ \emph {et~al.}(2015)\citenamefont {Younas}, \citenamefont {Hussain}, \citenamefont {Jamil},\ and\ \citenamefont {Bahamonde}}]{Younas:2015sva}%
  \BibitemOpen
  \bibfield  {author} {\bibinfo {author} {\bibfnamefont {A.}~\bibnamefont {Younas}}, \bibinfo {author} {\bibfnamefont {S.}~\bibnamefont {Hussain}}, \bibinfo {author} {\bibfnamefont {M.}~\bibnamefont {Jamil}}, \ and\ \bibinfo {author} {\bibfnamefont {S.}~\bibnamefont {Bahamonde}},\ }\href {\doibase 10.1103/PhysRevD.92.084042} {\bibfield  {journal} {\bibinfo  {journal} {Phys. Rev. D}\ }\textbf {\bibinfo {volume} {92}},\ \bibinfo {pages} {084042} (\bibinfo {year} {2015})},\ \Eprint {http://arxiv.org/abs/1502.01676} {arXiv:1502.01676 [gr-qc]} \BibitemShut {NoStop}%
\bibitem [{\citenamefont {Azreg-A{\"\i}nou}\ \emph {et~al.}(2017)\citenamefont {Azreg-A{\"\i}nou}, \citenamefont {Bahamonde},\ and\ \citenamefont {Jamil}}]{Azreg-Ainou:2017obt}%
  \BibitemOpen
  \bibfield  {author} {\bibinfo {author} {\bibfnamefont {M.}~\bibnamefont {Azreg-A{\"\i}nou}}, \bibinfo {author} {\bibfnamefont {S.}~\bibnamefont {Bahamonde}}, \ and\ \bibinfo {author} {\bibfnamefont {M.}~\bibnamefont {Jamil}},\ }\href {\doibase 10.1140/epjc/s10052-017-4969-4} {\bibfield  {journal} {\bibinfo  {journal} {Eur. Phys. J. C}\ }\textbf {\bibinfo {volume} {77}},\ \bibinfo {pages} {414} (\bibinfo {year} {2017})},\ \Eprint {http://arxiv.org/abs/1701.02239} {arXiv:1701.02239 [gr-qc]} \BibitemShut {NoStop}%
\bibitem [{\citenamefont {Zhu}\ \emph {et~al.}(2019)\citenamefont {Zhu}, \citenamefont {Wu}, \citenamefont {Jamil},\ and\ \citenamefont {Jusufi}}]{Zhu:2019ura}%
  \BibitemOpen
  \bibfield  {author} {\bibinfo {author} {\bibfnamefont {T.}~\bibnamefont {Zhu}}, \bibinfo {author} {\bibfnamefont {Q.}~\bibnamefont {Wu}}, \bibinfo {author} {\bibfnamefont {M.}~\bibnamefont {Jamil}}, \ and\ \bibinfo {author} {\bibfnamefont {K.}~\bibnamefont {Jusufi}},\ }\href {\doibase 10.1103/PhysRevD.100.044055} {\bibfield  {journal} {\bibinfo  {journal} {Phys. Rev. D}\ }\textbf {\bibinfo {volume} {100}},\ \bibinfo {pages} {044055} (\bibinfo {year} {2019})},\ \Eprint {http://arxiv.org/abs/1906.05673} {arXiv:1906.05673 [gr-qc]} \BibitemShut {NoStop}%
\bibitem [{\citenamefont {Wei}\ \emph {et~al.}(2015)\citenamefont {Wei}, \citenamefont {Yang},\ and\ \citenamefont {Liu}}]{Wei:2014dka}%
  \BibitemOpen
  \bibfield  {author} {\bibinfo {author} {\bibfnamefont {S.-W.}\ \bibnamefont {Wei}}, \bibinfo {author} {\bibfnamefont {K.}~\bibnamefont {Yang}}, \ and\ \bibinfo {author} {\bibfnamefont {Y.-X.}\ \bibnamefont {Liu}},\ }\href {\doibase 10.1140/epjc/s10052-015-3556-9} {\bibfield  {journal} {\bibinfo  {journal} {Eur. Phys. J. C}\ }\textbf {\bibinfo {volume} {75}},\ \bibinfo {pages} {253} (\bibinfo {year} {2015})},\ \bibinfo {note} {[Erratum: Eur.Phys.J.C 75, 331 (2015)]},\ \Eprint {http://arxiv.org/abs/1405.2178} {arXiv:1405.2178 [gr-qc]} \BibitemShut {NoStop}%
\bibitem [{\citenamefont {Sotani}\ and\ \citenamefont {Miyamoto}(2015)}]{Sotani:2015ewa}%
  \BibitemOpen
  \bibfield  {author} {\bibinfo {author} {\bibfnamefont {H.}~\bibnamefont {Sotani}}\ and\ \bibinfo {author} {\bibfnamefont {U.}~\bibnamefont {Miyamoto}},\ }\href {\doibase 10.1103/PhysRevD.92.044052} {\bibfield  {journal} {\bibinfo  {journal} {Phys. Rev. D}\ }\textbf {\bibinfo {volume} {92}},\ \bibinfo {pages} {044052} (\bibinfo {year} {2015})},\ \Eprint {http://arxiv.org/abs/1508.03119} {arXiv:1508.03119 [gr-qc]} \BibitemShut {NoStop}%
\bibitem [{\citenamefont {Babar}\ \emph {et~al.}(2021)\citenamefont {Babar}, \citenamefont {Atamurotov}, \citenamefont {Ul~Islam},\ and\ \citenamefont {Ghosh}}]{Babar:2021nst}%
  \BibitemOpen
  \bibfield  {author} {\bibinfo {author} {\bibfnamefont {G.~Z.}\ \bibnamefont {Babar}}, \bibinfo {author} {\bibfnamefont {F.}~\bibnamefont {Atamurotov}}, \bibinfo {author} {\bibfnamefont {S.}~\bibnamefont {Ul~Islam}}, \ and\ \bibinfo {author} {\bibfnamefont {S.~G.}\ \bibnamefont {Ghosh}},\ }\href {\doibase 10.1103/PhysRevD.103.084057} {\bibfield  {journal} {\bibinfo  {journal} {Phys. Rev. D}\ }\textbf {\bibinfo {volume} {103}},\ \bibinfo {pages} {084057} (\bibinfo {year} {2021})},\ \Eprint {http://arxiv.org/abs/2104.00714} {arXiv:2104.00714 [gr-qc]} \BibitemShut {NoStop}%
\bibitem [{\citenamefont {Islam}\ \emph {et~al.}(2020)\citenamefont {Islam}, \citenamefont {Kumar},\ and\ \citenamefont {Ghosh}}]{Islam:2020xmy}%
  \BibitemOpen
  \bibfield  {author} {\bibinfo {author} {\bibfnamefont {S.~U.}\ \bibnamefont {Islam}}, \bibinfo {author} {\bibfnamefont {R.}~\bibnamefont {Kumar}}, \ and\ \bibinfo {author} {\bibfnamefont {S.~G.}\ \bibnamefont {Ghosh}},\ }\href {\doibase 10.1088/1475-7516/2020/09/030} {\bibfield  {journal} {\bibinfo  {journal} {JCAP}\ }\textbf {\bibinfo {volume} {09}},\ \bibinfo {pages} {030} (\bibinfo {year} {2020})},\ \Eprint {http://arxiv.org/abs/2004.01038} {arXiv:2004.01038 [gr-qc]} \BibitemShut {NoStop}%
\bibitem [{\citenamefont {Kumar}\ \emph {et~al.}(2020)\citenamefont {Kumar}, \citenamefont {Islam},\ and\ \citenamefont {Ghosh}}]{Kumar:2020sag}%
  \BibitemOpen
  \bibfield  {author} {\bibinfo {author} {\bibfnamefont {R.}~\bibnamefont {Kumar}}, \bibinfo {author} {\bibfnamefont {S.~U.}\ \bibnamefont {Islam}}, \ and\ \bibinfo {author} {\bibfnamefont {S.~G.}\ \bibnamefont {Ghosh}},\ }\href {\doibase 10.1140/epjc/s10052-020-08606-3} {\bibfield  {journal} {\bibinfo  {journal} {Eur. Phys. J. C}\ }\textbf {\bibinfo {volume} {80}},\ \bibinfo {pages} {1128} (\bibinfo {year} {2020})},\ \Eprint {http://arxiv.org/abs/2004.12970} {arXiv:2004.12970 [gr-qc]} \BibitemShut {NoStop}%
\bibitem [{\citenamefont {Islam}\ \emph {et~al.}(2024)\citenamefont {Islam}, \citenamefont {Ghosh},\ and\ \citenamefont {Maharaj}}]{Islam:2022ybr}%
  \BibitemOpen
  \bibfield  {author} {\bibinfo {author} {\bibfnamefont {S.~U.}\ \bibnamefont {Islam}}, \bibinfo {author} {\bibfnamefont {S.~G.}\ \bibnamefont {Ghosh}}, \ and\ \bibinfo {author} {\bibfnamefont {S.~D.}\ \bibnamefont {Maharaj}},\ }\href {\doibase 10.1016/j.cjph.2024.03.044} {\bibfield  {journal} {\bibinfo  {journal} {Chin. J. Phys.}\ }\textbf {\bibinfo {volume} {89}},\ \bibinfo {pages} {1710} (\bibinfo {year} {2024})},\ \Eprint {http://arxiv.org/abs/2203.00957} {arXiv:2203.00957 [gr-qc]} \BibitemShut {NoStop}%
\bibitem [{\citenamefont {Kumar}\ \emph {et~al.}(2022{\natexlab{a}})\citenamefont {Kumar}, \citenamefont {Islam},\ and\ \citenamefont {Ghosh}}]{Kumar:2021cyl}%
  \BibitemOpen
  \bibfield  {author} {\bibinfo {author} {\bibfnamefont {J.}~\bibnamefont {Kumar}}, \bibinfo {author} {\bibfnamefont {S.~U.}\ \bibnamefont {Islam}}, \ and\ \bibinfo {author} {\bibfnamefont {S.~G.}\ \bibnamefont {Ghosh}},\ }\href {\doibase 10.1140/epjc/s10052-022-10357-2} {\bibfield  {journal} {\bibinfo  {journal} {Eur. Phys. J. C}\ }\textbf {\bibinfo {volume} {82}},\ \bibinfo {pages} {443} (\bibinfo {year} {2022}{\natexlab{a}})},\ \Eprint {http://arxiv.org/abs/2109.04450} {arXiv:2109.04450 [gr-qc]} \BibitemShut {NoStop}%
\bibitem [{\citenamefont {Walia}\ \emph {et~al.}(2022)\citenamefont {Walia}, \citenamefont {Maharaj},\ and\ \citenamefont {Ghosh}}]{Walia:2021emv}%
  \BibitemOpen
  \bibfield  {author} {\bibinfo {author} {\bibfnamefont {R.~K.}\ \bibnamefont {Walia}}, \bibinfo {author} {\bibfnamefont {S.~D.}\ \bibnamefont {Maharaj}}, \ and\ \bibinfo {author} {\bibfnamefont {S.~G.}\ \bibnamefont {Ghosh}},\ }\href {\doibase 10.1140/epjc/s10052-022-10451-5} {\bibfield  {journal} {\bibinfo  {journal} {Eur. Phys. J. C}\ }\textbf {\bibinfo {volume} {82}},\ \bibinfo {pages} {547} (\bibinfo {year} {2022})},\ \Eprint {http://arxiv.org/abs/2109.08055} {arXiv:2109.08055 [gr-qc]} \BibitemShut {NoStop}%
\bibitem [{\citenamefont {Islam}\ \emph {et~al.}(2021)\citenamefont {Islam}, \citenamefont {Kumar},\ and\ \citenamefont {Ghosh}}]{Islam:2021ful}%
  \BibitemOpen
  \bibfield  {author} {\bibinfo {author} {\bibfnamefont {S.~U.}\ \bibnamefont {Islam}}, \bibinfo {author} {\bibfnamefont {J.}~\bibnamefont {Kumar}}, \ and\ \bibinfo {author} {\bibfnamefont {S.~G.}\ \bibnamefont {Ghosh}},\ }\href {\doibase 10.1088/1475-7516/2021/10/013} {\bibfield  {journal} {\bibinfo  {journal} {JCAP}\ }\textbf {\bibinfo {volume} {10}},\ \bibinfo {pages} {013} (\bibinfo {year} {2021})},\ \Eprint {http://arxiv.org/abs/2104.00696} {arXiv:2104.00696 [gr-qc]} \BibitemShut {NoStop}%
\bibitem [{\citenamefont {Ghosh}\ and\ \citenamefont {Islam}(2023)}]{Ghosh:2023usx}%
  \BibitemOpen
  \bibfield  {author} {\bibinfo {author} {\bibfnamefont {S.~G.}\ \bibnamefont {Ghosh}}\ and\ \bibinfo {author} {\bibfnamefont {S.~U.}\ \bibnamefont {Islam}},\ }in\ \href {\doibase 10.1142/9789811269776_0316} {\emph {\bibinfo {booktitle} {{16th Marcel Grossmann Meeting on~Recent Developments in Theoretical and Experimental General Relativity, Astrophysics and Relativistic Field Theories}}}}\ (\bibinfo {year} {2023})\BibitemShut {NoStop}%
\bibitem [{\citenamefont {Vachher}\ \emph {et~al.}(2025{\natexlab{a}})\citenamefont {Vachher}, \citenamefont {Kumar},\ and\ \citenamefont {Ghosh}}]{Vachher:2025jsq}%
  \BibitemOpen
  \bibfield  {author} {\bibinfo {author} {\bibfnamefont {A.}~\bibnamefont {Vachher}}, \bibinfo {author} {\bibfnamefont {A.}~\bibnamefont {Kumar}}, \ and\ \bibinfo {author} {\bibfnamefont {S.~G.}\ \bibnamefont {Ghosh}},\ }\href {\doibase 10.1088/1475-7516/2025/11/021} {\bibfield  {journal} {\bibinfo  {journal} {JCAP}\ }\textbf {\bibinfo {volume} {11}},\ \bibinfo {pages} {021} (\bibinfo {year} {2025}{\natexlab{a}})},\ \Eprint {http://arxiv.org/abs/2508.21100} {arXiv:2508.21100 [gr-qc]} \BibitemShut {NoStop}%
\bibitem [{\citenamefont {Ghosh}\ \emph {et~al.}(2021)\citenamefont {Ghosh}, \citenamefont {Kumar},\ and\ \citenamefont {Islam}}]{Ghosh:2020spb}%
  \BibitemOpen
  \bibfield  {author} {\bibinfo {author} {\bibfnamefont {S.~G.}\ \bibnamefont {Ghosh}}, \bibinfo {author} {\bibfnamefont {R.}~\bibnamefont {Kumar}}, \ and\ \bibinfo {author} {\bibfnamefont {S.~U.}\ \bibnamefont {Islam}},\ }\href {\doibase 10.1088/1475-7516/2021/03/056} {\bibfield  {journal} {\bibinfo  {journal} {JCAP}\ }\textbf {\bibinfo {volume} {03}},\ \bibinfo {pages} {056} (\bibinfo {year} {2021})},\ \Eprint {http://arxiv.org/abs/2011.08023} {arXiv:2011.08023 [gr-qc]} \BibitemShut {NoStop}%
\bibitem [{\citenamefont {Liu}\ \emph {et~al.}(2017)\citenamefont {Liu}, \citenamefont {Ding},\ and\ \citenamefont {Jing}}]{Liu:2016eju}%
  \BibitemOpen
  \bibfield  {author} {\bibinfo {author} {\bibfnamefont {C.-Q.}\ \bibnamefont {Liu}}, \bibinfo {author} {\bibfnamefont {C.-K.}\ \bibnamefont {Ding}}, \ and\ \bibinfo {author} {\bibfnamefont {J.-L.}\ \bibnamefont {Jing}},\ }\href {\doibase 10.1088/0256-307X/34/9/090401} {\bibfield  {journal} {\bibinfo  {journal} {Chin. Phys. Lett.}\ }\textbf {\bibinfo {volume} {34}},\ \bibinfo {pages} {090401} (\bibinfo {year} {2017})},\ \Eprint {http://arxiv.org/abs/1610.02128} {arXiv:1610.02128 [gr-qc]} \BibitemShut {NoStop}%
\bibitem [{\citenamefont {Bisnovatyi-Kogan}\ and\ \citenamefont {Tsupko}(2017)}]{Bisnovatyi-Kogan:2017kii}%
  \BibitemOpen
  \bibfield  {author} {\bibinfo {author} {\bibfnamefont {G.~S.}\ \bibnamefont {Bisnovatyi-Kogan}}\ and\ \bibinfo {author} {\bibfnamefont {O.~Y.}\ \bibnamefont {Tsupko}},\ }\href {\doibase 10.3390/universe3030057} {\bibfield  {journal} {\bibinfo  {journal} {Universe}\ }\textbf {\bibinfo {volume} {3}},\ \bibinfo {pages} {57} (\bibinfo {year} {2017})},\ \Eprint {http://arxiv.org/abs/1905.06615} {arXiv:1905.06615 [gr-qc]} \BibitemShut {NoStop}%
\bibitem [{\citenamefont {Feleppa}\ \emph {et~al.}(2024)\citenamefont {Feleppa}, \citenamefont {Bozza},\ and\ \citenamefont {Tsupko}}]{Feleppa:2024vdk}%
  \BibitemOpen
  \bibfield  {author} {\bibinfo {author} {\bibfnamefont {F.}~\bibnamefont {Feleppa}}, \bibinfo {author} {\bibfnamefont {V.}~\bibnamefont {Bozza}}, \ and\ \bibinfo {author} {\bibfnamefont {O.~Y.}\ \bibnamefont {Tsupko}},\ }\href {\doibase 10.1103/PhysRevD.110.064031} {\bibfield  {journal} {\bibinfo  {journal} {Phys. Rev. D}\ }\textbf {\bibinfo {volume} {110}},\ \bibinfo {pages} {064031} (\bibinfo {year} {2024})},\ \Eprint {http://arxiv.org/abs/2406.07703} {arXiv:2406.07703 [gr-qc]} \BibitemShut {NoStop}%
\bibitem [{\citenamefont {Turakhonov}\ \emph {et~al.}(2025)\citenamefont {Turakhonov}, \citenamefont {Atamurotov}, \citenamefont {Ghosh},\ and\ \citenamefont {Abdujabbarov}}]{Turakhonov:2025ojy}%
  \BibitemOpen
  \bibfield  {author} {\bibinfo {author} {\bibfnamefont {Z.}~\bibnamefont {Turakhonov}}, \bibinfo {author} {\bibfnamefont {F.}~\bibnamefont {Atamurotov}}, \bibinfo {author} {\bibfnamefont {S.~G.}\ \bibnamefont {Ghosh}}, \ and\ \bibinfo {author} {\bibfnamefont {A.}~\bibnamefont {Abdujabbarov}},\ }\href {\doibase 10.1016/j.dark.2025.101880} {\bibfield  {journal} {\bibinfo  {journal} {Phys. Dark Univ.}\ }\textbf {\bibinfo {volume} {48}},\ \bibinfo {pages} {101880} (\bibinfo {year} {2025})}\BibitemShut {NoStop}%
\bibitem [{\citenamefont {Umarov}\ \emph {et~al.}(2025)\citenamefont {Umarov}, \citenamefont {Yunusov}, \citenamefont {Atamurotov}, \citenamefont {Abdujabbarov},\ and\ \citenamefont {Ghosh}}]{Umarov:2025btg}%
  \BibitemOpen
  \bibfield  {author} {\bibinfo {author} {\bibfnamefont {D.}~\bibnamefont {Umarov}}, \bibinfo {author} {\bibfnamefont {O.}~\bibnamefont {Yunusov}}, \bibinfo {author} {\bibfnamefont {F.}~\bibnamefont {Atamurotov}}, \bibinfo {author} {\bibfnamefont {A.}~\bibnamefont {Abdujabbarov}}, \ and\ \bibinfo {author} {\bibfnamefont {S.~G.}\ \bibnamefont {Ghosh}},\ }\href {\doibase 10.1088/1674-1137/adb384} {\bibfield  {journal} {\bibinfo  {journal} {Chin. Phys. C}\ }\textbf {\bibinfo {volume} {49}},\ \bibinfo {pages} {055102} (\bibinfo {year} {2025})}\BibitemShut {NoStop}%
\bibitem [{\citenamefont {Stefanov}\ \emph {et~al.}(2010)\citenamefont {Stefanov}, \citenamefont {Yazadjiev},\ and\ \citenamefont {Gyulchev}}]{Stefanov:2010xz}%
  \BibitemOpen
  \bibfield  {author} {\bibinfo {author} {\bibfnamefont {I.~Z.}\ \bibnamefont {Stefanov}}, \bibinfo {author} {\bibfnamefont {S.~S.}\ \bibnamefont {Yazadjiev}}, \ and\ \bibinfo {author} {\bibfnamefont {G.~G.}\ \bibnamefont {Gyulchev}},\ }\href {\doibase 10.1103/PhysRevLett.104.251103} {\bibfield  {journal} {\bibinfo  {journal} {Phys. Rev. Lett.}\ }\textbf {\bibinfo {volume} {104}},\ \bibinfo {pages} {251103} (\bibinfo {year} {2010})},\ \Eprint {http://arxiv.org/abs/1003.1609} {arXiv:1003.1609 [gr-qc]} \BibitemShut {NoStop}%
\bibitem [{\citenamefont {Raffaelli}(2016)}]{Raffaelli:2014ola}%
  \BibitemOpen
  \bibfield  {author} {\bibinfo {author} {\bibfnamefont {B.}~\bibnamefont {Raffaelli}},\ }\href {\doibase 10.1007/s10714-016-2016-7} {\bibfield  {journal} {\bibinfo  {journal} {Gen. Rel. Grav.}\ }\textbf {\bibinfo {volume} {48}},\ \bibinfo {pages} {16} (\bibinfo {year} {2016})},\ \Eprint {http://arxiv.org/abs/1412.7333} {arXiv:1412.7333 [gr-qc]} \BibitemShut {NoStop}%
\bibitem [{\citenamefont {Virbhadra}(2022)}]{Virbhadra:2022iiy}%
  \BibitemOpen
  \bibfield  {author} {\bibinfo {author} {\bibfnamefont {K.~S.}\ \bibnamefont {Virbhadra}},\ }\href {\doibase 10.1103/PhysRevD.106.064038} {\bibfield  {journal} {\bibinfo  {journal} {Phys. Rev. D}\ }\textbf {\bibinfo {volume} {106}},\ \bibinfo {pages} {064038} (\bibinfo {year} {2022})},\ \Eprint {http://arxiv.org/abs/2204.01879} {arXiv:2204.01879 [gr-qc]} \BibitemShut {NoStop}%
\bibitem [{\citenamefont {Virbhadra}(2024)}]{Virbhadra:2024xpk}%
  \BibitemOpen
  \bibfield  {author} {\bibinfo {author} {\bibfnamefont {K.~S.}\ \bibnamefont {Virbhadra}},\ }\href {\doibase 10.1103/PhysRevD.109.124004} {\bibfield  {journal} {\bibinfo  {journal} {Phys. Rev. D}\ }\textbf {\bibinfo {volume} {109}},\ \bibinfo {pages} {124004} (\bibinfo {year} {2024})},\ \Eprint {http://arxiv.org/abs/2402.17190} {arXiv:2402.17190 [gr-qc]} \BibitemShut {NoStop}%
\bibitem [{\citenamefont {Adler}\ and\ \citenamefont {Virbhadra}(2022)}]{Adler:2022qtb}%
  \BibitemOpen
  \bibfield  {author} {\bibinfo {author} {\bibfnamefont {S.~L.}\ \bibnamefont {Adler}}\ and\ \bibinfo {author} {\bibfnamefont {K.~S.}\ \bibnamefont {Virbhadra}},\ }\href {\doibase 10.1007/s10714-022-02976-7} {\bibfield  {journal} {\bibinfo  {journal} {Gen. Rel. Grav.}\ }\textbf {\bibinfo {volume} {54}},\ \bibinfo {pages} {93} (\bibinfo {year} {2022})},\ \Eprint {http://arxiv.org/abs/2205.04628} {arXiv:2205.04628 [gr-qc]} \BibitemShut {NoStop}%
\bibitem [{\citenamefont {Kuang}\ and\ \citenamefont {\"Ovg\"un}(2022)}]{Kuang:2022xjp}%
  \BibitemOpen
  \bibfield  {author} {\bibinfo {author} {\bibfnamefont {X.-M.}\ \bibnamefont {Kuang}}\ and\ \bibinfo {author} {\bibfnamefont {A.}~\bibnamefont {\"Ovg\"un}},\ }\href {\doibase 10.1016/j.aop.2022.169147} {\bibfield  {journal} {\bibinfo  {journal} {Annals Phys.}\ }\textbf {\bibinfo {volume} {447}},\ \bibinfo {pages} {169147} (\bibinfo {year} {2022})},\ \Eprint {http://arxiv.org/abs/2205.11003} {arXiv:2205.11003 [gr-qc]} \BibitemShut {NoStop}%
\bibitem [{\citenamefont {Kuang}\ \emph {et~al.}(2022)\citenamefont {Kuang}, \citenamefont {Tang}, \citenamefont {Wang},\ and\ \citenamefont {Wang}}]{Kuang:2022ojj}%
  \BibitemOpen
  \bibfield  {author} {\bibinfo {author} {\bibfnamefont {X.-M.}\ \bibnamefont {Kuang}}, \bibinfo {author} {\bibfnamefont {Z.-Y.}\ \bibnamefont {Tang}}, \bibinfo {author} {\bibfnamefont {B.}~\bibnamefont {Wang}}, \ and\ \bibinfo {author} {\bibfnamefont {A.}~\bibnamefont {Wang}},\ }\href {\doibase 10.1103/PhysRevD.106.064012} {\bibfield  {journal} {\bibinfo  {journal} {Phys. Rev. D}\ }\textbf {\bibinfo {volume} {106}},\ \bibinfo {pages} {064012} (\bibinfo {year} {2022})},\ \Eprint {http://arxiv.org/abs/2206.05878} {arXiv:2206.05878 [gr-qc]} \BibitemShut {NoStop}%
\bibitem [{\citenamefont {Kumar}\ \emph {et~al.}(2023)\citenamefont {Kumar}, \citenamefont {Islam},\ and\ \citenamefont {Ghosh}}]{Kumar:2023jgh}%
  \BibitemOpen
  \bibfield  {author} {\bibinfo {author} {\bibfnamefont {J.}~\bibnamefont {Kumar}}, \bibinfo {author} {\bibfnamefont {S.~U.}\ \bibnamefont {Islam}}, \ and\ \bibinfo {author} {\bibfnamefont {S.~G.}\ \bibnamefont {Ghosh}},\ }\href {\doibase 10.1140/epjc/s10052-023-12205-3} {\bibfield  {journal} {\bibinfo  {journal} {Eur. Phys. J. C}\ }\textbf {\bibinfo {volume} {83}},\ \bibinfo {pages} {1014} (\bibinfo {year} {2023})},\ \Eprint {http://arxiv.org/abs/2305.04336} {arXiv:2305.04336 [gr-qc]} \BibitemShut {NoStop}%
\bibitem [{\citenamefont {Molla}\ \emph {et~al.}(2024{\natexlab{a}})\citenamefont {Molla}, \citenamefont {Chaudhary}, \citenamefont {Mustafa}, \citenamefont {Debnath},\ and\ \citenamefont {Maurya}}]{Molla:2024yde}%
  \BibitemOpen
  \bibfield  {author} {\bibinfo {author} {\bibfnamefont {N.~U.}\ \bibnamefont {Molla}}, \bibinfo {author} {\bibfnamefont {H.}~\bibnamefont {Chaudhary}}, \bibinfo {author} {\bibfnamefont {G.}~\bibnamefont {Mustafa}}, \bibinfo {author} {\bibfnamefont {U.}~\bibnamefont {Debnath}}, \ and\ \bibinfo {author} {\bibfnamefont {S.~K.}\ \bibnamefont {Maurya}},\ }\href {\doibase 10.1140/epjc/s10052-024-12679-9} {\bibfield  {journal} {\bibinfo  {journal} {Eur. Phys. J. C}\ }\textbf {\bibinfo {volume} {84}},\ \bibinfo {pages} {390} (\bibinfo {year} {2024}{\natexlab{a}})}\BibitemShut {NoStop}%
\bibitem [{\citenamefont {Zhao}\ \emph {et~al.}(2025)\citenamefont {Zhao}, \citenamefont {Tang},\ and\ \citenamefont {Xu}}]{Zhao:2024hep}%
  \BibitemOpen
  \bibfield  {author} {\bibinfo {author} {\bibfnamefont {L.}~\bibnamefont {Zhao}}, \bibinfo {author} {\bibfnamefont {M.}~\bibnamefont {Tang}}, \ and\ \bibinfo {author} {\bibfnamefont {Z.}~\bibnamefont {Xu}},\ }\href {\doibase 10.1140/epjc/s10052-025-14190-1} {\bibfield  {journal} {\bibinfo  {journal} {Eur. Phys. J. C}\ }\textbf {\bibinfo {volume} {85}},\ \bibinfo {pages} {452} (\bibinfo {year} {2025})},\ \Eprint {http://arxiv.org/abs/2408.01205} {arXiv:2408.01205 [gr-qc]} \BibitemShut {NoStop}%
\bibitem [{\citenamefont {Vachher}\ and\ \citenamefont {Ghosh}(2025)}]{Vachher:2024ait}%
  \BibitemOpen
  \bibfield  {author} {\bibinfo {author} {\bibfnamefont {A.}~\bibnamefont {Vachher}}\ and\ \bibinfo {author} {\bibfnamefont {S.~G.}\ \bibnamefont {Ghosh}},\ }\href {\doibase 10.1016/j.jheap.2024.11.012} {\bibfield  {journal} {\bibinfo  {journal} {JHEAp}\ }\textbf {\bibinfo {volume} {45}},\ \bibinfo {pages} {75} (\bibinfo {year} {2025})},\ \Eprint {http://arxiv.org/abs/2410.11332} {arXiv:2410.11332 [gr-qc]} \BibitemShut {NoStop}%
\bibitem [{\citenamefont {Kumar}\ \emph {et~al.}(2022{\natexlab{b}})\citenamefont {Kumar}, \citenamefont {Islam},\ and\ \citenamefont {Ghosh}}]{Kumar:2022fqo}%
  \BibitemOpen
  \bibfield  {author} {\bibinfo {author} {\bibfnamefont {J.}~\bibnamefont {Kumar}}, \bibinfo {author} {\bibfnamefont {S.~U.}\ \bibnamefont {Islam}}, \ and\ \bibinfo {author} {\bibfnamefont {S.~G.}\ \bibnamefont {Ghosh}},\ }\href {\doibase 10.3847/1538-4357/ac912c} {\bibfield  {journal} {\bibinfo  {journal} {Astrophys. J.}\ }\textbf {\bibinfo {volume} {938}},\ \bibinfo {pages} {104} (\bibinfo {year} {2022}{\natexlab{b}})},\ \Eprint {http://arxiv.org/abs/2209.04240} {arXiv:2209.04240 [gr-qc]} \BibitemShut {NoStop}%
\bibitem [{\citenamefont {Molla}\ \emph {et~al.}(2024{\natexlab{b}})\citenamefont {Molla}, \citenamefont {Ghosh},\ and\ \citenamefont {Debnath}}]{Molla:2024lpt}%
  \BibitemOpen
  \bibfield  {author} {\bibinfo {author} {\bibfnamefont {N.~U.}\ \bibnamefont {Molla}}, \bibinfo {author} {\bibfnamefont {S.~G.}\ \bibnamefont {Ghosh}}, \ and\ \bibinfo {author} {\bibfnamefont {U.}~\bibnamefont {Debnath}},\ }\href {\doibase 10.1016/j.dark.2024.101495} {\bibfield  {journal} {\bibinfo  {journal} {Phys. Dark Univ.}\ }\textbf {\bibinfo {volume} {44}},\ \bibinfo {pages} {101495} (\bibinfo {year} {2024}{\natexlab{b}})}\BibitemShut {NoStop}%
\bibitem [{\citenamefont {Vachher}\ \emph {et~al.}(2025{\natexlab{b}})\citenamefont {Vachher}, \citenamefont {Islam}, \citenamefont {Kumar~Walia},\ and\ \citenamefont {Ghosh}}]{Vachher:2024ezs}%
  \BibitemOpen
  \bibfield  {author} {\bibinfo {author} {\bibfnamefont {A.}~\bibnamefont {Vachher}}, \bibinfo {author} {\bibfnamefont {S.~U.}\ \bibnamefont {Islam}}, \bibinfo {author} {\bibfnamefont {R.}~\bibnamefont {Kumar~Walia}}, \ and\ \bibinfo {author} {\bibfnamefont {S.~G.}\ \bibnamefont {Ghosh}},\ }\href@noop {} {\bibfield  {journal} {\bibinfo  {journal} {Annals Phys.}\ }\textbf {\bibinfo {volume} {480}},\ \bibinfo {pages} {170084} (\bibinfo {year} {2025}{\natexlab{b}})},\ \Eprint {http://arxiv.org/abs/2405.06501} {arXiv:2405.06501 [gr-qc]} \BibitemShut {NoStop}%
\bibitem [{\citenamefont {Ali}\ \emph {et~al.}(2025{\natexlab{a}})\citenamefont {Ali}, \citenamefont {Islam}, \citenamefont {Ghosh},\ and\ \citenamefont {Ramasamya}}]{Ali:2024mrt}%
  \BibitemOpen
  \bibfield  {author} {\bibinfo {author} {\bibfnamefont {A.}~\bibnamefont {Ali}}, \bibinfo {author} {\bibfnamefont {S.~U.}\ \bibnamefont {Islam}}, \bibinfo {author} {\bibfnamefont {S.~G.}\ \bibnamefont {Ghosh}}, \ and\ \bibinfo {author} {\bibfnamefont {A.}~\bibnamefont {Ramasamya}},\ }\href {\doibase 10.1016/j.dark.2024.101768} {\bibfield  {journal} {\bibinfo  {journal} {Phys. Dark Univ.}\ }\textbf {\bibinfo {volume} {47}},\ \bibinfo {pages} {101768} (\bibinfo {year} {2025}{\natexlab{a}})}\BibitemShut {NoStop}%
\bibitem [{\citenamefont {Yan}\ \emph {et~al.}(2025)\citenamefont {Yan}, \citenamefont {Zhu},\ and\ \citenamefont {Wu}}]{Yan:2025mlg}%
  \BibitemOpen
  \bibfield  {author} {\bibinfo {author} {\bibfnamefont {J.-M.}\ \bibnamefont {Yan}}, \bibinfo {author} {\bibfnamefont {T.}~\bibnamefont {Zhu}}, \ and\ \bibinfo {author} {\bibfnamefont {Q.}~\bibnamefont {Wu}},\ }\href@noop {} {\  (\bibinfo {year} {2025})},\ \Eprint {http://arxiv.org/abs/2504.10956} {arXiv:2504.10956 [gr-qc]} \BibitemShut {NoStop}%
\bibitem [{\citenamefont {Shi}\ and\ \citenamefont {Zhu}(2024)}]{Shi:2024bpm}%
  \BibitemOpen
  \bibfield  {author} {\bibinfo {author} {\bibfnamefont {H.-Y.}\ \bibnamefont {Shi}}\ and\ \bibinfo {author} {\bibfnamefont {T.}~\bibnamefont {Zhu}},\ }\href {\doibase 10.1140/epjc/s10052-024-13198-3} {\bibfield  {journal} {\bibinfo  {journal} {Eur. Phys. J. C}\ }\textbf {\bibinfo {volume} {84}},\ \bibinfo {pages} {814} (\bibinfo {year} {2024})}\BibitemShut {NoStop}%
\bibitem [{\citenamefont {Vachher}\ \emph {et~al.}(2024)\citenamefont {Vachher}, \citenamefont {Baboolal},\ and\ \citenamefont {Ghosh}}]{Vachher:2024ldc}%
  \BibitemOpen
  \bibfield  {author} {\bibinfo {author} {\bibfnamefont {A.}~\bibnamefont {Vachher}}, \bibinfo {author} {\bibfnamefont {D.}~\bibnamefont {Baboolal}}, \ and\ \bibinfo {author} {\bibfnamefont {S.~G.}\ \bibnamefont {Ghosh}},\ }\href {\doibase 10.1016/j.dark.2024.101493} {\bibfield  {journal} {\bibinfo  {journal} {Phys. Dark Univ.}\ }\textbf {\bibinfo {volume} {44}},\ \bibinfo {pages} {101493} (\bibinfo {year} {2024})}\BibitemShut {NoStop}%
\bibitem [{\citenamefont {Ali}\ \emph {et~al.}(2025{\natexlab{b}})\citenamefont {Ali}, \citenamefont {Molla}, \citenamefont {Ghosh}, \citenamefont {Ramasamya},\ and\ \citenamefont {Debnath}}]{Ali:2025rop}%
  \BibitemOpen
  \bibfield  {author} {\bibinfo {author} {\bibfnamefont {A.}~\bibnamefont {Ali}}, \bibinfo {author} {\bibfnamefont {N.~U.}\ \bibnamefont {Molla}}, \bibinfo {author} {\bibfnamefont {S.~G.}\ \bibnamefont {Ghosh}}, \bibinfo {author} {\bibfnamefont {A.}~\bibnamefont {Ramasamya}}, \ and\ \bibinfo {author} {\bibfnamefont {U.}~\bibnamefont {Debnath}},\ }\href {\doibase 10.1016/j.dark.2025.101859} {\bibfield  {journal} {\bibinfo  {journal} {Phys. Dark Univ.}\ }\textbf {\bibinfo {volume} {48}},\ \bibinfo {pages} {101859} (\bibinfo {year} {2025}{\natexlab{b}})}\BibitemShut {NoStop}%
\bibitem [{\citenamefont {Ali}\ \emph {et~al.}(2025{\natexlab{c}})\citenamefont {Ali}, \citenamefont {Ramasamya},\ and\ \citenamefont {Ghosh}}]{Ali:2025ney}%
  \BibitemOpen
  \bibfield  {author} {\bibinfo {author} {\bibfnamefont {A.}~\bibnamefont {Ali}}, \bibinfo {author} {\bibfnamefont {A.}~\bibnamefont {Ramasamya}}, \ and\ \bibinfo {author} {\bibfnamefont {S.~G.}\ \bibnamefont {Ghosh}},\ }\href {\doibase 10.1016/j.jheap.2025.100418} {\bibfield  {journal} {\bibinfo  {journal} {JHEAp}\ }\textbf {\bibinfo {volume} {48}},\ \bibinfo {pages} {100418} (\bibinfo {year} {2025}{\natexlab{c}})}\BibitemShut {NoStop}%
\bibitem [{\citenamefont {Jusufi}\ \emph {et~al.}(2019)\citenamefont {Jusufi}, \citenamefont {Jamil}, \citenamefont {Salucci}, \citenamefont {Zhu},\ and\ \citenamefont {Haroon}}]{Jusufi:2019nrn}%
  \BibitemOpen
  \bibfield  {author} {\bibinfo {author} {\bibfnamefont {K.}~\bibnamefont {Jusufi}}, \bibinfo {author} {\bibfnamefont {M.}~\bibnamefont {Jamil}}, \bibinfo {author} {\bibfnamefont {P.}~\bibnamefont {Salucci}}, \bibinfo {author} {\bibfnamefont {T.}~\bibnamefont {Zhu}}, \ and\ \bibinfo {author} {\bibfnamefont {S.}~\bibnamefont {Haroon}},\ }\href {\doibase 10.1103/PhysRevD.100.044012} {\bibfield  {journal} {\bibinfo  {journal} {Phys. Rev. D}\ }\textbf {\bibinfo {volume} {100}},\ \bibinfo {pages} {044012} (\bibinfo {year} {2019})},\ \Eprint {http://arxiv.org/abs/1905.11803} {arXiv:1905.11803 [physics.gen-ph]} \BibitemShut {NoStop}%
\bibitem [{\citenamefont {Jusufi}\ \emph {et~al.}(2020)\citenamefont {Jusufi}, \citenamefont {Jamil},\ and\ \citenamefont {Zhu}}]{Jusufi:2020cpn}%
  \BibitemOpen
  \bibfield  {author} {\bibinfo {author} {\bibfnamefont {K.}~\bibnamefont {Jusufi}}, \bibinfo {author} {\bibfnamefont {M.}~\bibnamefont {Jamil}}, \ and\ \bibinfo {author} {\bibfnamefont {T.}~\bibnamefont {Zhu}},\ }\href {\doibase 10.1140/epjc/s10052-020-7899-5} {\bibfield  {journal} {\bibinfo  {journal} {Eur. Phys. J. C}\ }\textbf {\bibinfo {volume} {80}},\ \bibinfo {pages} {354} (\bibinfo {year} {2020})},\ \Eprint {http://arxiv.org/abs/2005.05299} {arXiv:2005.05299 [gr-qc]} \BibitemShut {NoStop}%
\bibitem [{\citenamefont {Modesto}(2004)}]{Modesto:2004wm}%
  \BibitemOpen
  \bibfield  {author} {\bibinfo {author} {\bibfnamefont {L.}~\bibnamefont {Modesto}},\ }\href {\doibase 10.1103/PhysRevD.70.124009} {\bibfield  {journal} {\bibinfo  {journal} {Phys. Rev. D}\ }\textbf {\bibinfo {volume} {70}},\ \bibinfo {pages} {124009} (\bibinfo {year} {2004})},\ \Eprint {http://arxiv.org/abs/gr-qc/0407097} {arXiv:gr-qc/0407097} \BibitemShut {NoStop}%
\bibitem [{\citenamefont {Modesto}(2006)}]{Modesto:2005zm}%
  \BibitemOpen
  \bibfield  {author} {\bibinfo {author} {\bibfnamefont {L.}~\bibnamefont {Modesto}},\ }\href {\doibase 10.1103/PhysRevD.73.104004} {\bibfield  {journal} {\bibinfo  {journal} {Phys. Rev. D}\ }\textbf {\bibinfo {volume} {73}},\ \bibinfo {pages} {104004} (\bibinfo {year} {2006})},\ \Eprint {http://arxiv.org/abs/gr-qc/0509078} {arXiv:gr-qc/0509078} \BibitemShut {NoStop}%
\bibitem [{\citenamefont {Modesto}(2010)}]{Modesto:2008im}%
  \BibitemOpen
  \bibfield  {author} {\bibinfo {author} {\bibfnamefont {L.}~\bibnamefont {Modesto}},\ }\href {\doibase 10.1007/s10773-010-0346-x} {\bibfield  {journal} {\bibinfo  {journal} {Int. J. Theor. Phys.}\ }\textbf {\bibinfo {volume} {49}},\ \bibinfo {pages} {1649} (\bibinfo {year} {2010})},\ \Eprint {http://arxiv.org/abs/0811.2196} {arXiv:0811.2196 [gr-qc]} \BibitemShut {NoStop}%
\bibitem [{\citenamefont {Ashtekar}\ \emph {et~al.}(2018)\citenamefont {Ashtekar}, \citenamefont {Olmedo},\ and\ \citenamefont {Singh}}]{Ashtekar:2018lag}%
  \BibitemOpen
  \bibfield  {author} {\bibinfo {author} {\bibfnamefont {A.}~\bibnamefont {Ashtekar}}, \bibinfo {author} {\bibfnamefont {J.}~\bibnamefont {Olmedo}}, \ and\ \bibinfo {author} {\bibfnamefont {P.}~\bibnamefont {Singh}},\ }\href {\doibase 10.1103/PhysRevD.98.126003} {\bibfield  {journal} {\bibinfo  {journal} {Phys. Rev. D}\ }\textbf {\bibinfo {volume} {98}},\ \bibinfo {pages} {126003} (\bibinfo {year} {2018})},\ \Eprint {http://arxiv.org/abs/1806.00648} {arXiv:1806.00648 [gr-qc]} \BibitemShut {NoStop}%
\bibitem [{\citenamefont {Ashtekar}\ \emph {et~al.}(2020)\citenamefont {Ashtekar}, \citenamefont {Olmedo},\ and\ \citenamefont {Singh}}]{Ashtekar:2020ckv}%
  \BibitemOpen
  \bibfield  {author} {\bibinfo {author} {\bibfnamefont {A.}~\bibnamefont {Ashtekar}}, \bibinfo {author} {\bibfnamefont {J.}~\bibnamefont {Olmedo}}, \ and\ \bibinfo {author} {\bibfnamefont {P.}~\bibnamefont {Singh}},\ }\href {\doibase 10.1088/1361-6382/ab7245} {\bibfield  {journal} {\bibinfo  {journal} {Class. Quant. Grav.}\ }\textbf {\bibinfo {volume} {37}},\ \bibinfo {pages} {205017} (\bibinfo {year} {2020})},\ \Eprint {http://arxiv.org/abs/2001.08833} {arXiv:2001.08833 [gr-qc]} \BibitemShut {NoStop}%
\bibitem [{\citenamefont {Bojowald}\ and\ \citenamefont {Modesto}(2006)}]{Bojowald:2005qw}%
  \BibitemOpen
  \bibfield  {author} {\bibinfo {author} {\bibfnamefont {M.}~\bibnamefont {Bojowald}}\ and\ \bibinfo {author} {\bibfnamefont {L.}~\bibnamefont {Modesto}},\ }\href {\doibase 10.1088/0264-9381/23/2/010} {\bibfield  {journal} {\bibinfo  {journal} {Class. Quant. Grav.}\ }\textbf {\bibinfo {volume} {23}},\ \bibinfo {pages} {3429} (\bibinfo {year} {2006})},\ \Eprint {http://arxiv.org/abs/gr-qc/0509075} {arXiv:gr-qc/0509075} \BibitemShut {NoStop}%
\bibitem [{\citenamefont {Chiou}(2008)}]{Chiou:2008eg}%
  \BibitemOpen
  \bibfield  {author} {\bibinfo {author} {\bibfnamefont {D.-W.}\ \bibnamefont {Chiou}},\ }\href {\doibase 10.1103/PhysRevD.78.064040} {\bibfield  {journal} {\bibinfo  {journal} {Phys. Rev. D}\ }\textbf {\bibinfo {volume} {78}},\ \bibinfo {pages} {064040} (\bibinfo {year} {2008})},\ \Eprint {http://arxiv.org/abs/0807.0665} {arXiv:0807.0665 [gr-qc]} \BibitemShut {NoStop}%
\bibitem [{\citenamefont {Jiang}\ \emph {et~al.}(2024{\natexlab{a}})\citenamefont {Jiang}, \citenamefont {Liu}, \citenamefont {Dihingia}, \citenamefont {Mizuno}, \citenamefont {Xu}, \citenamefont {Zhu},\ and\ \citenamefont {Wu}}]{Jiang:2023img}%
  \BibitemOpen
  \bibfield  {author} {\bibinfo {author} {\bibfnamefont {H.-X.}\ \bibnamefont {Jiang}}, \bibinfo {author} {\bibfnamefont {C.}~\bibnamefont {Liu}}, \bibinfo {author} {\bibfnamefont {I.~K.}\ \bibnamefont {Dihingia}}, \bibinfo {author} {\bibfnamefont {Y.}~\bibnamefont {Mizuno}}, \bibinfo {author} {\bibfnamefont {H.}~\bibnamefont {Xu}}, \bibinfo {author} {\bibfnamefont {T.}~\bibnamefont {Zhu}}, \ and\ \bibinfo {author} {\bibfnamefont {Q.}~\bibnamefont {Wu}},\ }\href {\doibase 10.1088/1475-7516/2024/01/059} {\bibfield  {journal} {\bibinfo  {journal} {JCAP}\ }\textbf {\bibinfo {volume} {01}},\ \bibinfo {pages} {059} (\bibinfo {year} {2024}{\natexlab{a}})},\ \Eprint {http://arxiv.org/abs/2312.04288} {arXiv:2312.04288 [gr-qc]} \BibitemShut {NoStop}%
\bibitem [{\citenamefont {Liu}\ \emph {et~al.}(2023)\citenamefont {Liu}, \citenamefont {Siew}, \citenamefont {Zhu}, \citenamefont {Wu}, \citenamefont {Sun}, \citenamefont {Zhao},\ and\ \citenamefont {Xu}}]{Liu:2023vfh}%
  \BibitemOpen
  \bibfield  {author} {\bibinfo {author} {\bibfnamefont {C.}~\bibnamefont {Liu}}, \bibinfo {author} {\bibfnamefont {H.}~\bibnamefont {Siew}}, \bibinfo {author} {\bibfnamefont {T.}~\bibnamefont {Zhu}}, \bibinfo {author} {\bibfnamefont {Q.}~\bibnamefont {Wu}}, \bibinfo {author} {\bibfnamefont {Y.}~\bibnamefont {Sun}}, \bibinfo {author} {\bibfnamefont {Y.}~\bibnamefont {Zhao}}, \ and\ \bibinfo {author} {\bibfnamefont {H.}~\bibnamefont {Xu}},\ }\href {\doibase 10.1088/1475-7516/2023/11/096} {\bibfield  {journal} {\bibinfo  {journal} {JCAP}\ }\textbf {\bibinfo {volume} {11}},\ \bibinfo {pages} {096} (\bibinfo {year} {2023})},\ \Eprint {http://arxiv.org/abs/2305.12323} {arXiv:2305.12323 [gr-qc]} \BibitemShut {NoStop}%
\bibitem [{\citenamefont {Tu}\ \emph {et~al.}(2023)\citenamefont {Tu}, \citenamefont {Zhu},\ and\ \citenamefont {Wang}}]{Tu:2023xab}%
  \BibitemOpen
  \bibfield  {author} {\bibinfo {author} {\bibfnamefont {Z.-Y.}\ \bibnamefont {Tu}}, \bibinfo {author} {\bibfnamefont {T.}~\bibnamefont {Zhu}}, \ and\ \bibinfo {author} {\bibfnamefont {A.}~\bibnamefont {Wang}},\ }\href {\doibase 10.1103/PhysRevD.108.024035} {\bibfield  {journal} {\bibinfo  {journal} {Phys. Rev. D}\ }\textbf {\bibinfo {volume} {108}},\ \bibinfo {pages} {024035} (\bibinfo {year} {2023})},\ \Eprint {http://arxiv.org/abs/2304.14160} {arXiv:2304.14160 [gr-qc]} \BibitemShut {NoStop}%
\bibitem [{\citenamefont {Liu}\ \emph {et~al.}(2020)\citenamefont {Liu}, \citenamefont {Zhu}, \citenamefont {Wu}, \citenamefont {Jusufi}, \citenamefont {Jamil}, \citenamefont {Azreg-A{\"\i}nou},\ and\ \citenamefont {Wang}}]{Liu:2020ola}%
  \BibitemOpen
  \bibfield  {author} {\bibinfo {author} {\bibfnamefont {C.}~\bibnamefont {Liu}}, \bibinfo {author} {\bibfnamefont {T.}~\bibnamefont {Zhu}}, \bibinfo {author} {\bibfnamefont {Q.}~\bibnamefont {Wu}}, \bibinfo {author} {\bibfnamefont {K.}~\bibnamefont {Jusufi}}, \bibinfo {author} {\bibfnamefont {M.}~\bibnamefont {Jamil}}, \bibinfo {author} {\bibfnamefont {M.}~\bibnamefont {Azreg-A{\"\i}nou}}, \ and\ \bibinfo {author} {\bibfnamefont {A.}~\bibnamefont {Wang}},\ }\href {\doibase 10.1103/PhysRevD.101.084001} {\bibfield  {journal} {\bibinfo  {journal} {Phys. Rev. D}\ }\textbf {\bibinfo {volume} {101}},\ \bibinfo {pages} {084001} (\bibinfo {year} {2020})},\ \bibinfo {note} {[Erratum: Phys.Rev.D 103, 089902 (2021)]},\ \Eprint {http://arxiv.org/abs/2003.00477} {arXiv:2003.00477 [gr-qc]} \BibitemShut {NoStop}%
\bibitem [{\citenamefont {Zhu}\ and\ \citenamefont {Wang}(2020)}]{Zhu:2020tcf}%
  \BibitemOpen
  \bibfield  {author} {\bibinfo {author} {\bibfnamefont {T.}~\bibnamefont {Zhu}}\ and\ \bibinfo {author} {\bibfnamefont {A.}~\bibnamefont {Wang}},\ }\href {\doibase 10.1103/PhysRevD.102.124042} {\bibfield  {journal} {\bibinfo  {journal} {Phys. Rev. D}\ }\textbf {\bibinfo {volume} {102}},\ \bibinfo {pages} {124042} (\bibinfo {year} {2020})},\ \Eprint {http://arxiv.org/abs/2008.08704} {arXiv:2008.08704 [gr-qc]} \BibitemShut {NoStop}%
\bibitem [{\citenamefont {Yan}\ \emph {et~al.}(2022)\citenamefont {Yan}, \citenamefont {Wu}, \citenamefont {Liu}, \citenamefont {Zhu},\ and\ \citenamefont {Wang}}]{Yan:2022fkr}%
  \BibitemOpen
  \bibfield  {author} {\bibinfo {author} {\bibfnamefont {J.-M.}\ \bibnamefont {Yan}}, \bibinfo {author} {\bibfnamefont {Q.}~\bibnamefont {Wu}}, \bibinfo {author} {\bibfnamefont {C.}~\bibnamefont {Liu}}, \bibinfo {author} {\bibfnamefont {T.}~\bibnamefont {Zhu}}, \ and\ \bibinfo {author} {\bibfnamefont {A.}~\bibnamefont {Wang}},\ }\href {\doibase 10.1088/1475-7516/2022/09/008} {\bibfield  {journal} {\bibinfo  {journal} {JCAP}\ }\textbf {\bibinfo {volume} {09}},\ \bibinfo {pages} {008} (\bibinfo {year} {2022})},\ \Eprint {http://arxiv.org/abs/2203.03203} {arXiv:2203.03203 [gr-qc]} \BibitemShut {NoStop}%
\bibitem [{\citenamefont {Yan}\ \emph {et~al.}(2023)\citenamefont {Yan}, \citenamefont {Liu}, \citenamefont {Zhu}, \citenamefont {Wu},\ and\ \citenamefont {Wang}}]{Yan:2023vdg}%
  \BibitemOpen
  \bibfield  {author} {\bibinfo {author} {\bibfnamefont {J.-M.}\ \bibnamefont {Yan}}, \bibinfo {author} {\bibfnamefont {C.}~\bibnamefont {Liu}}, \bibinfo {author} {\bibfnamefont {T.}~\bibnamefont {Zhu}}, \bibinfo {author} {\bibfnamefont {Q.}~\bibnamefont {Wu}}, \ and\ \bibinfo {author} {\bibfnamefont {A.}~\bibnamefont {Wang}},\ }\href {\doibase 10.1103/PhysRevD.107.084043} {\bibfield  {journal} {\bibinfo  {journal} {Phys. Rev. D}\ }\textbf {\bibinfo {volume} {107}},\ \bibinfo {pages} {084043} (\bibinfo {year} {2023})},\ \Eprint {http://arxiv.org/abs/2302.10482} {arXiv:2302.10482 [gr-qc]} \BibitemShut {NoStop}%
\bibitem [{\citenamefont {Jiang}\ \emph {et~al.}(2024{\natexlab{b}})\citenamefont {Jiang}, \citenamefont {Dihingia}, \citenamefont {Liu}, \citenamefont {Mizuno},\ and\ \citenamefont {Zhu}}]{Jiang:2024vgn}%
  \BibitemOpen
  \bibfield  {author} {\bibinfo {author} {\bibfnamefont {H.-X.}\ \bibnamefont {Jiang}}, \bibinfo {author} {\bibfnamefont {I.~K.}\ \bibnamefont {Dihingia}}, \bibinfo {author} {\bibfnamefont {C.}~\bibnamefont {Liu}}, \bibinfo {author} {\bibfnamefont {Y.}~\bibnamefont {Mizuno}}, \ and\ \bibinfo {author} {\bibfnamefont {T.}~\bibnamefont {Zhu}},\ }\href {\doibase 10.1088/1475-7516/2024/05/101} {\bibfield  {journal} {\bibinfo  {journal} {JCAP}\ }\textbf {\bibinfo {volume} {05}},\ \bibinfo {pages} {101} (\bibinfo {year} {2024}{\natexlab{b}})},\ \Eprint {http://arxiv.org/abs/2402.08402} {arXiv:2402.08402 [astro-ph.HE]} \BibitemShut {NoStop}%
\bibitem [{\citenamefont {Jiang}\ \emph {et~al.}(2024{\natexlab{c}})\citenamefont {Jiang}, \citenamefont {Alloqulov}, \citenamefont {Wu}, \citenamefont {Shaymatov},\ and\ \citenamefont {Zhu}}]{Jiang:2024cpe}%
  \BibitemOpen
  \bibfield  {author} {\bibinfo {author} {\bibfnamefont {H.}~\bibnamefont {Jiang}}, \bibinfo {author} {\bibfnamefont {M.}~\bibnamefont {Alloqulov}}, \bibinfo {author} {\bibfnamefont {Q.}~\bibnamefont {Wu}}, \bibinfo {author} {\bibfnamefont {S.}~\bibnamefont {Shaymatov}}, \ and\ \bibinfo {author} {\bibfnamefont {T.}~\bibnamefont {Zhu}},\ }\href {\doibase 10.1016/j.dark.2024.101627} {\bibfield  {journal} {\bibinfo  {journal} {Phys. Dark Univ.}\ }\textbf {\bibinfo {volume} {46}},\ \bibinfo {pages} {101627} (\bibinfo {year} {2024}{\natexlab{c}})}\BibitemShut {NoStop}%
\bibitem [{\citenamefont {Uktamov}\ \emph {et~al.}(2025)\citenamefont {Uktamov}, \citenamefont {Alloqulov}, \citenamefont {Shaymatov}, \citenamefont {Zhu},\ and\ \citenamefont {Ahmedov}}]{Uktamov:2024ckf}%
  \BibitemOpen
  \bibfield  {author} {\bibinfo {author} {\bibfnamefont {U.}~\bibnamefont {Uktamov}}, \bibinfo {author} {\bibfnamefont {M.}~\bibnamefont {Alloqulov}}, \bibinfo {author} {\bibfnamefont {S.}~\bibnamefont {Shaymatov}}, \bibinfo {author} {\bibfnamefont {T.}~\bibnamefont {Zhu}}, \ and\ \bibinfo {author} {\bibfnamefont {B.}~\bibnamefont {Ahmedov}},\ }\href {\doibase 10.1016/j.dark.2024.101743} {\bibfield  {journal} {\bibinfo  {journal} {Phys. Dark Univ.}\ }\textbf {\bibinfo {volume} {47}},\ \bibinfo {pages} {101743} (\bibinfo {year} {2025})},\ \Eprint {http://arxiv.org/abs/2412.01809} {arXiv:2412.01809 [gr-qc]} \BibitemShut {NoStop}%
\bibitem [{\citenamefont {Xamidov}\ \emph {et~al.}(2025{\natexlab{a}})\citenamefont {Xamidov}, \citenamefont {Sheoran}, \citenamefont {Shaymatov},\ and\ \citenamefont {Zhu}}]{Xamidov:2024xpc}%
  \BibitemOpen
  \bibfield  {author} {\bibinfo {author} {\bibfnamefont {T.}~\bibnamefont {Xamidov}}, \bibinfo {author} {\bibfnamefont {P.}~\bibnamefont {Sheoran}}, \bibinfo {author} {\bibfnamefont {S.}~\bibnamefont {Shaymatov}}, \ and\ \bibinfo {author} {\bibfnamefont {T.}~\bibnamefont {Zhu}},\ }\href {\doibase 10.1088/1475-7516/2025/03/053} {\bibfield  {journal} {\bibinfo  {journal} {JCAP}\ }\textbf {\bibinfo {volume} {03}},\ \bibinfo {pages} {053} (\bibinfo {year} {2025}{\natexlab{a}})},\ \Eprint {http://arxiv.org/abs/2412.03885} {arXiv:2412.03885 [gr-qc]} \BibitemShut {NoStop}%
\bibitem [{\citenamefont {Xamidov}\ \emph {et~al.}(2025{\natexlab{b}})\citenamefont {Xamidov}, \citenamefont {Shaymatov}, \citenamefont {Ahmedov},\ and\ \citenamefont {Zhu}}]{Xamidov:2025oqx}%
  \BibitemOpen
  \bibfield  {author} {\bibinfo {author} {\bibfnamefont {T.}~\bibnamefont {Xamidov}}, \bibinfo {author} {\bibfnamefont {S.}~\bibnamefont {Shaymatov}}, \bibinfo {author} {\bibfnamefont {B.}~\bibnamefont {Ahmedov}}, \ and\ \bibinfo {author} {\bibfnamefont {T.}~\bibnamefont {Zhu}},\ }\href@noop {} {\  (\bibinfo {year} {2025}{\natexlab{b}})},\ \Eprint {http://arxiv.org/abs/2503.06750} {arXiv:2503.06750 [gr-qc]} \BibitemShut {NoStop}%
\bibitem [{\citenamefont {Borges}\ \emph {et~al.}(2024{\natexlab{b}})\citenamefont {Borges}, \citenamefont {Baranov}, \citenamefont {Sobrinho},\ and\ \citenamefont {Carneiro}}]{Borges:2023fub}%
  \BibitemOpen
  \bibfield  {author} {\bibinfo {author} {\bibfnamefont {H.~A.}\ \bibnamefont {Borges}}, \bibinfo {author} {\bibfnamefont {I.~P.~R.}\ \bibnamefont {Baranov}}, \bibinfo {author} {\bibfnamefont {F.~C.}\ \bibnamefont {Sobrinho}}, \ and\ \bibinfo {author} {\bibfnamefont {S.}~\bibnamefont {Carneiro}},\ }\href {\doibase 10.1088/1361-6382/ad210c} {\bibfield  {journal} {\bibinfo  {journal} {Class. Quant. Grav.}\ }\textbf {\bibinfo {volume} {41}},\ \bibinfo {pages} {05LT01} (\bibinfo {year} {2024}{\natexlab{b}})},\ \Eprint {http://arxiv.org/abs/2310.01560} {arXiv:2310.01560 [gr-qc]} \BibitemShut {NoStop}%
\bibitem [{\citenamefont {Wang}\ \emph {et~al.}(2025)\citenamefont {Wang}, \citenamefont {Vachher}, \citenamefont {Wu}, \citenamefont {Zhu},\ and\ \citenamefont {Ghosh}}]{Wang:2024iwt}%
  \BibitemOpen
  \bibfield  {author} {\bibinfo {author} {\bibfnamefont {Y.}~\bibnamefont {Wang}}, \bibinfo {author} {\bibfnamefont {A.}~\bibnamefont {Vachher}}, \bibinfo {author} {\bibfnamefont {Q.}~\bibnamefont {Wu}}, \bibinfo {author} {\bibfnamefont {T.}~\bibnamefont {Zhu}}, \ and\ \bibinfo {author} {\bibfnamefont {S.~G.}\ \bibnamefont {Ghosh}},\ }\href {\doibase 10.1140/epjc/s10052-025-13970-z} {\bibfield  {journal} {\bibinfo  {journal} {Eur. Phys. J. C}\ }\textbf {\bibinfo {volume} {85}},\ \bibinfo {pages} {302} (\bibinfo {year} {2025})},\ \Eprint {http://arxiv.org/abs/2410.12382} {arXiv:2410.12382 [astro-ph.CO]} \BibitemShut {NoStop}%
\bibitem [{\citenamefont {Tsukamoto}(2017)}]{Tsukamoto:2016jzh}%
  \BibitemOpen
  \bibfield  {author} {\bibinfo {author} {\bibfnamefont {N.}~\bibnamefont {Tsukamoto}},\ }\href {\doibase 10.1103/PhysRevD.95.064035} {\bibfield  {journal} {\bibinfo  {journal} {Phys. Rev. D}\ }\textbf {\bibinfo {volume} {95}},\ \bibinfo {pages} {064035} (\bibinfo {year} {2017})},\ \Eprint {http://arxiv.org/abs/1612.08251} {arXiv:1612.08251 [gr-qc]} \BibitemShut {NoStop}%
\bibitem [{\citenamefont {Bambi}\ \emph {et~al.}(2019)\citenamefont {Bambi}, \citenamefont {Freese}, \citenamefont {Vagnozzi},\ and\ \citenamefont {Visinelli}}]{Bambi:2019tjh}%
  \BibitemOpen
  \bibfield  {author} {\bibinfo {author} {\bibfnamefont {C.}~\bibnamefont {Bambi}}, \bibinfo {author} {\bibfnamefont {K.}~\bibnamefont {Freese}}, \bibinfo {author} {\bibfnamefont {S.}~\bibnamefont {Vagnozzi}}, \ and\ \bibinfo {author} {\bibfnamefont {L.}~\bibnamefont {Visinelli}},\ }\href {\doibase 10.1103/PhysRevD.100.044057} {\bibfield  {journal} {\bibinfo  {journal} {Phys. Rev. D}\ }\textbf {\bibinfo {volume} {100}},\ \bibinfo {pages} {044057} (\bibinfo {year} {2019})},\ \Eprint {http://arxiv.org/abs/1904.12983} {arXiv:1904.12983 [gr-qc]} \BibitemShut {NoStop}%
\bibitem [{\citenamefont {Cunha}\ and\ \citenamefont {Herdeiro}(2018)}]{Cunha:2018acu}%
  \BibitemOpen
  \bibfield  {author} {\bibinfo {author} {\bibfnamefont {P.~V.~P.}\ \bibnamefont {Cunha}}\ and\ \bibinfo {author} {\bibfnamefont {C.~A.~R.}\ \bibnamefont {Herdeiro}},\ }\href {\doibase 10.1007/s10714-018-2361-9} {\bibfield  {journal} {\bibinfo  {journal} {Gen. Rel. Grav.}\ }\textbf {\bibinfo {volume} {50}},\ \bibinfo {pages} {42} (\bibinfo {year} {2018})},\ \Eprint {http://arxiv.org/abs/1801.00860} {arXiv:1801.00860 [gr-qc]} \BibitemShut {NoStop}%
\bibitem [{\citenamefont {Psaltis}(2019)}]{Psaltis:2018xkc}%
  \BibitemOpen
  \bibfield  {author} {\bibinfo {author} {\bibfnamefont {D.}~\bibnamefont {Psaltis}},\ }\href {\doibase 10.1007/s10714-019-2611-5} {\bibfield  {journal} {\bibinfo  {journal} {Gen. Rel. Grav.}\ }\textbf {\bibinfo {volume} {51}},\ \bibinfo {pages} {137} (\bibinfo {year} {2019})},\ \Eprint {http://arxiv.org/abs/1806.09740} {arXiv:1806.09740 [astro-ph.HE]} \BibitemShut {NoStop}%
\bibitem [{\citenamefont {Vagnozzi}\ \emph {et~al.}(2023)\citenamefont {Vagnozzi} \emph {et~al.}}]{Vagnozzi:2022moj}%
  \BibitemOpen
  \bibfield  {author} {\bibinfo {author} {\bibfnamefont {S.}~\bibnamefont {Vagnozzi}} \emph {et~al.},\ }\href {\doibase 10.1088/1361-6382/acd97b} {\bibfield  {journal} {\bibinfo  {journal} {Class. Quant. Grav.}\ }\textbf {\bibinfo {volume} {40}},\ \bibinfo {pages} {165007} (\bibinfo {year} {2023})},\ \Eprint {http://arxiv.org/abs/2205.07787} {arXiv:2205.07787 [gr-qc]} \BibitemShut {NoStop}%
\bibitem [{\citenamefont {Afrin}\ \emph {et~al.}(2023)\citenamefont {Afrin}, \citenamefont {Vagnozzi},\ and\ \citenamefont {Ghosh}}]{Afrin:2022ztr}%
  \BibitemOpen
  \bibfield  {author} {\bibinfo {author} {\bibfnamefont {M.}~\bibnamefont {Afrin}}, \bibinfo {author} {\bibfnamefont {S.}~\bibnamefont {Vagnozzi}}, \ and\ \bibinfo {author} {\bibfnamefont {S.~G.}\ \bibnamefont {Ghosh}},\ }\href {\doibase 10.3847/1538-4357/acb334} {\bibfield  {journal} {\bibinfo  {journal} {Astrophys. J.}\ }\textbf {\bibinfo {volume} {944}},\ \bibinfo {pages} {149} (\bibinfo {year} {2023})},\ \Eprint {http://arxiv.org/abs/2209.12584} {arXiv:2209.12584 [gr-qc]} \BibitemShut {NoStop}%
\bibitem [{\citenamefont {Ghosh}\ \emph {et~al.}(2022)\citenamefont {Ghosh}, \citenamefont {Afrin},\ and\ \citenamefont {Kumar}}]{Ghosh:2022mka}%
  \BibitemOpen
  \bibfield  {author} {\bibinfo {author} {\bibfnamefont {S.~G.}\ \bibnamefont {Ghosh}}, \bibinfo {author} {\bibfnamefont {M.}~\bibnamefont {Afrin}}, \ and\ \bibinfo {author} {\bibfnamefont {R.}~\bibnamefont {Kumar}},\ }\href {\doibase 10.1103/PhysRevD.106.104013} {\bibfield  {journal} {\bibinfo  {journal} {Phys. Rev. D}\ }\textbf {\bibinfo {volume} {106}},\ \bibinfo {pages} {104013} (\bibinfo {year} {2022})},\ \Eprint {http://arxiv.org/abs/2206.04091} {arXiv:2206.04091 [gr-qc]} \BibitemShut {NoStop}%
\bibitem [{\citenamefont {Elizaga~Navascu{\'e}s}\ \emph {et~al.}(2022)\citenamefont {Elizaga~Navascu{\'e}s}, \citenamefont {Garc{\'\i}a-Quismondo},\ and\ \citenamefont {Mena~Marug{\'a}n}}]{ElizagaNavascues:2022rof}%
  \BibitemOpen
  \bibfield  {author} {\bibinfo {author} {\bibfnamefont {B.}~\bibnamefont {Elizaga~Navascu{\'e}s}}, \bibinfo {author} {\bibfnamefont {A.}~\bibnamefont {Garc{\'\i}a-Quismondo}}, \ and\ \bibinfo {author} {\bibfnamefont {G.~A.}\ \bibnamefont {Mena~Marug{\'a}n}},\ }\href {\doibase 10.1103/PhysRevD.106.043531} {\bibfield  {journal} {\bibinfo  {journal} {Phys. Rev. D}\ }\textbf {\bibinfo {volume} {106}},\ \bibinfo {pages} {043531} (\bibinfo {year} {2022})},\ \Eprint {http://arxiv.org/abs/2208.00425} {arXiv:2208.00425 [gr-qc]} \BibitemShut {NoStop}%
\bibitem [{\citenamefont {Alonso-Bardaji}\ \emph {et~al.}(2022)\citenamefont {Alonso-Bardaji}, \citenamefont {Brizuela},\ and\ \citenamefont {Vera}}]{Alonso-Bardaji:2021yls}%
  \BibitemOpen
  \bibfield  {author} {\bibinfo {author} {\bibfnamefont {A.}~\bibnamefont {Alonso-Bardaji}}, \bibinfo {author} {\bibfnamefont {D.}~\bibnamefont {Brizuela}}, \ and\ \bibinfo {author} {\bibfnamefont {R.}~\bibnamefont {Vera}},\ }\href {\doibase 10.1016/j.physletb.2022.137075} {\bibfield  {journal} {\bibinfo  {journal} {Phys. Lett. B}\ }\textbf {\bibinfo {volume} {829}},\ \bibinfo {pages} {137075} (\bibinfo {year} {2022})},\ \Eprint {http://arxiv.org/abs/2112.12110} {arXiv:2112.12110 [gr-qc]} \BibitemShut {NoStop}%
\bibitem [{\citenamefont {Alonso-Bardaji}\ \emph {et~al.}(2023)\citenamefont {Alonso-Bardaji}, \citenamefont {Brizuela},\ and\ \citenamefont {Vera}}]{Alonso-Bardaji:2023niu}%
  \BibitemOpen
  \bibfield  {author} {\bibinfo {author} {\bibfnamefont {A.}~\bibnamefont {Alonso-Bardaji}}, \bibinfo {author} {\bibfnamefont {D.}~\bibnamefont {Brizuela}}, \ and\ \bibinfo {author} {\bibfnamefont {R.}~\bibnamefont {Vera}},\ }\href {\doibase 10.1103/PhysRevD.107.064067} {\bibfield  {journal} {\bibinfo  {journal} {Phys. Rev. D}\ }\textbf {\bibinfo {volume} {107}},\ \bibinfo {pages} {064067} (\bibinfo {year} {2023})},\ \Eprint {http://arxiv.org/abs/2302.10619} {arXiv:2302.10619 [gr-qc]} \BibitemShut {NoStop}%
\bibitem [{\citenamefont {Tibrewala}(2012)}]{Tibrewala:2012xb}%
  \BibitemOpen
  \bibfield  {author} {\bibinfo {author} {\bibfnamefont {R.}~\bibnamefont {Tibrewala}},\ }\href {\doibase 10.1088/0264-9381/29/23/235012} {\bibfield  {journal} {\bibinfo  {journal} {Class. Quant. Grav.}\ }\textbf {\bibinfo {volume} {29}},\ \bibinfo {pages} {235012} (\bibinfo {year} {2012})},\ \Eprint {http://arxiv.org/abs/1207.2585} {arXiv:1207.2585 [gr-qc]} \BibitemShut {NoStop}%
\bibitem [{\citenamefont {Gambini}\ \emph {et~al.}(2015)\citenamefont {Gambini}, \citenamefont {Capurro},\ and\ \citenamefont {Pullin}}]{Gambini:2014qta}%
  \BibitemOpen
  \bibfield  {author} {\bibinfo {author} {\bibfnamefont {R.}~\bibnamefont {Gambini}}, \bibinfo {author} {\bibfnamefont {E.~M.}\ \bibnamefont {Capurro}}, \ and\ \bibinfo {author} {\bibfnamefont {J.}~\bibnamefont {Pullin}},\ }\href {\doibase 10.1103/PhysRevD.91.084006} {\bibfield  {journal} {\bibinfo  {journal} {Phys. Rev. D}\ }\textbf {\bibinfo {volume} {91}},\ \bibinfo {pages} {084006} (\bibinfo {year} {2015})},\ \Eprint {http://arxiv.org/abs/1412.6055} {arXiv:1412.6055 [gr-qc]} \BibitemShut {NoStop}%
\bibitem [{\citenamefont {Gambini}\ \emph {et~al.}(2022)\citenamefont {Gambini}, \citenamefont {Ben{\'\i}tez},\ and\ \citenamefont {Pullin}}]{Gambini:2021uzf}%
  \BibitemOpen
  \bibfield  {author} {\bibinfo {author} {\bibfnamefont {R.}~\bibnamefont {Gambini}}, \bibinfo {author} {\bibfnamefont {F.}~\bibnamefont {Ben{\'\i}tez}}, \ and\ \bibinfo {author} {\bibfnamefont {J.}~\bibnamefont {Pullin}},\ }\href {\doibase 10.3390/universe8100526} {\bibfield  {journal} {\bibinfo  {journal} {Universe}\ }\textbf {\bibinfo {volume} {8}},\ \bibinfo {pages} {526} (\bibinfo {year} {2022})},\ \Eprint {http://arxiv.org/abs/2102.09501} {arXiv:2102.09501 [gr-qc]} \BibitemShut {NoStop}%
\bibitem [{\citenamefont {Domagala}\ and\ \citenamefont {Lewandowski}(2004)}]{Domagala:2004jt}%
  \BibitemOpen
  \bibfield  {author} {\bibinfo {author} {\bibfnamefont {M.}~\bibnamefont {Domagala}}\ and\ \bibinfo {author} {\bibfnamefont {J.}~\bibnamefont {Lewandowski}},\ }\href {\doibase 10.1088/0264-9381/21/22/014} {\bibfield  {journal} {\bibinfo  {journal} {Class. Quant. Grav.}\ }\textbf {\bibinfo {volume} {21}},\ \bibinfo {pages} {5233} (\bibinfo {year} {2004})},\ \Eprint {http://arxiv.org/abs/gr-qc/0407051} {arXiv:gr-qc/0407051} \BibitemShut {NoStop}%
\bibitem [{\citenamefont {Meissner}(2004)}]{Meissner:2004ju}%
  \BibitemOpen
  \bibfield  {author} {\bibinfo {author} {\bibfnamefont {K.~A.}\ \bibnamefont {Meissner}},\ }\href {\doibase 10.1088/0264-9381/21/22/015} {\bibfield  {journal} {\bibinfo  {journal} {Class. Quant. Grav.}\ }\textbf {\bibinfo {volume} {21}},\ \bibinfo {pages} {5245} (\bibinfo {year} {2004})},\ \Eprint {http://arxiv.org/abs/gr-qc/0407052} {arXiv:gr-qc/0407052} \BibitemShut {NoStop}%
\bibitem [{\citenamefont {Corichi}\ \emph {et~al.}(2007)\citenamefont {Corichi}, \citenamefont {Diaz-Polo},\ and\ \citenamefont {Fernandez-Borja}}]{Corichi:2006bs}%
  \BibitemOpen
  \bibfield  {author} {\bibinfo {author} {\bibfnamefont {A.}~\bibnamefont {Corichi}}, \bibinfo {author} {\bibfnamefont {J.}~\bibnamefont {Diaz-Polo}}, \ and\ \bibinfo {author} {\bibfnamefont {E.}~\bibnamefont {Fernandez-Borja}},\ }\href {\doibase 10.1088/0264-9381/24/1/013} {\bibfield  {journal} {\bibinfo  {journal} {Class. Quant. Grav.}\ }\textbf {\bibinfo {volume} {24}},\ \bibinfo {pages} {243} (\bibinfo {year} {2007})},\ \Eprint {http://arxiv.org/abs/gr-qc/0605014} {arXiv:gr-qc/0605014} \BibitemShut {NoStop}%
\bibitem [{\citenamefont {Pigozzo}\ \emph {et~al.}(2021)\citenamefont {Pigozzo}, \citenamefont {Bacelar},\ and\ \citenamefont {Carneiro}}]{Pigozzo:2020zft}%
  \BibitemOpen
  \bibfield  {author} {\bibinfo {author} {\bibfnamefont {C.}~\bibnamefont {Pigozzo}}, \bibinfo {author} {\bibfnamefont {F.~S.}\ \bibnamefont {Bacelar}}, \ and\ \bibinfo {author} {\bibfnamefont {S.}~\bibnamefont {Carneiro}},\ }\href {\doibase 10.1088/1361-6382/abce6a} {\bibfield  {journal} {\bibinfo  {journal} {Class. Quant. Grav.}\ }\textbf {\bibinfo {volume} {38}},\ \bibinfo {pages} {045001} (\bibinfo {year} {2021})},\ \Eprint {http://arxiv.org/abs/2001.03440} {arXiv:2001.03440 [gr-qc]} \BibitemShut {NoStop}%
\bibitem [{\citenamefont {Chandrasekhar}(1985)}]{Chandrasekhar:1985kt}%
  \BibitemOpen
  \bibfield  {author} {\bibinfo {author} {\bibfnamefont {S.}~\bibnamefont {Chandrasekhar}},\ }\href@noop {} {\emph {\bibinfo {title} {{The mathematical theory of black holes}}}}\ (\bibinfo {year} {1985})\BibitemShut {NoStop}%
\bibitem [{\citenamefont {Kumar}\ and\ \citenamefont {Ghosh}(2026)}]{KUMAR2026170291}%
  \BibitemOpen
  \bibfield  {author} {\bibinfo {author} {\bibfnamefont {A.}~\bibnamefont {Kumar}}\ and\ \bibinfo {author} {\bibfnamefont {S.~G.}\ \bibnamefont {Ghosh}},\ }\href {\doibase https://doi.org/10.1016/j.aop.2025.170291} {\bibfield  {journal} {\bibinfo  {journal} {Annals of Physics}\ }\textbf {\bibinfo {volume} {484}},\ \bibinfo {pages} {170291} (\bibinfo {year} {2026})}\BibitemShut {NoStop}%
\bibitem [{\citenamefont {Berti}\ \emph {et~al.}(2015)\citenamefont {Berti} \emph {et~al.}}]{Berti:2015itd}%
  \BibitemOpen
  \bibfield  {author} {\bibinfo {author} {\bibfnamefont {E.}~\bibnamefont {Berti}} \emph {et~al.},\ }\href {\doibase 10.1088/0264-9381/32/24/243001} {\bibfield  {journal} {\bibinfo  {journal} {Class. Quant. Grav.}\ }\textbf {\bibinfo {volume} {32}},\ \bibinfo {pages} {243001} (\bibinfo {year} {2015})},\ \Eprint {http://arxiv.org/abs/1501.07274} {arXiv:1501.07274 [gr-qc]} \BibitemShut {NoStop}%
\bibitem [{\citenamefont {Cardoso}\ and\ \citenamefont {Pani}(2019)}]{Cardoso:2019rvt}%
  \BibitemOpen
  \bibfield  {author} {\bibinfo {author} {\bibfnamefont {V.}~\bibnamefont {Cardoso}}\ and\ \bibinfo {author} {\bibfnamefont {P.}~\bibnamefont {Pani}},\ }\href {\doibase 10.1007/s41114-019-0020-4} {\bibfield  {journal} {\bibinfo  {journal} {Living Rev. Rel.}\ }\textbf {\bibinfo {volume} {22}},\ \bibinfo {pages} {4} (\bibinfo {year} {2019})},\ \Eprint {http://arxiv.org/abs/1904.05363} {arXiv:1904.05363 [gr-qc]} \BibitemShut {NoStop}%
\bibitem [{\citenamefont {Kumar~Walia}(2023)}]{KumarWalia:2022ddq}%
  \BibitemOpen
  \bibfield  {author} {\bibinfo {author} {\bibfnamefont {R.}~\bibnamefont {Kumar~Walia}},\ }\href {\doibase 10.1088/1475-7516/2023/03/029} {\bibfield  {journal} {\bibinfo  {journal} {JCAP}\ }\textbf {\bibinfo {volume} {03}},\ \bibinfo {pages} {029} (\bibinfo {year} {2023})},\ \Eprint {http://arxiv.org/abs/2207.02106} {arXiv:2207.02106 [gr-qc]} \BibitemShut {NoStop}%
\bibitem [{\citenamefont {Kumar}\ \emph {et~al.}(2025)\citenamefont {Kumar}, \citenamefont {Islam},\ and\ \citenamefont {Ghosh}}]{Kumar:2025bim}%
  \BibitemOpen
  \bibfield  {author} {\bibinfo {author} {\bibfnamefont {A.}~\bibnamefont {Kumar}}, \bibinfo {author} {\bibfnamefont {S.~U.}\ \bibnamefont {Islam}}, \ and\ \bibinfo {author} {\bibfnamefont {S.~G.}\ \bibnamefont {Ghosh}},\ }\href@noop {} {\  (\bibinfo {year} {2025})},\ \Eprint {http://arxiv.org/abs/2509.00127} {arXiv:2509.00127 [gr-qc]} \BibitemShut {NoStop}%
\bibitem [{\citenamefont {Smith}\ \emph {et~al.}(2005)\citenamefont {Smith}, \citenamefont {Blakeslee}, \citenamefont {Lucey},\ and\ \citenamefont {Tonry}}]{Smith:2005pq}%
  \BibitemOpen
  \bibfield  {author} {\bibinfo {author} {\bibfnamefont {R.~J.}\ \bibnamefont {Smith}}, \bibinfo {author} {\bibfnamefont {J.~P.}\ \bibnamefont {Blakeslee}}, \bibinfo {author} {\bibfnamefont {J.~R.}\ \bibnamefont {Lucey}}, \ and\ \bibinfo {author} {\bibfnamefont {J.}~\bibnamefont {Tonry}},\ }\href {\doibase 10.1086/431240} {\bibfield  {journal} {\bibinfo  {journal} {Astrophys. J. Lett.}\ }\textbf {\bibinfo {volume} {625}},\ \bibinfo {pages} {L103} (\bibinfo {year} {2005})},\ \Eprint {http://arxiv.org/abs/astro-ph/0504453} {arXiv:astro-ph/0504453} \BibitemShut {NoStop}%
\bibitem [{\citenamefont {Smith}\ and\ \citenamefont {Lucey}(2013)}]{Smith:2013ena}%
  \BibitemOpen
  \bibfield  {author} {\bibinfo {author} {\bibfnamefont {R.~J.}\ \bibnamefont {Smith}}\ and\ \bibinfo {author} {\bibfnamefont {J.~R.}\ \bibnamefont {Lucey}},\ }\href {\doibase 10.1093/mnras/stt1141} {\bibfield  {journal} {\bibinfo  {journal} {Mon. Not. Roy. Astron. Soc.}\ }\textbf {\bibinfo {volume} {434}},\ \bibinfo {pages} {1964} (\bibinfo {year} {2013})},\ \Eprint {http://arxiv.org/abs/1306.4983} {arXiv:1306.4983 [astro-ph.CO]} \BibitemShut {NoStop}%
\bibitem [{\citenamefont {Bozza}(2008)}]{Bozza:2008ev}%
  \BibitemOpen
  \bibfield  {author} {\bibinfo {author} {\bibfnamefont {V.}~\bibnamefont {Bozza}},\ }\href {\doibase 10.1103/PhysRevD.78.103005} {\bibfield  {journal} {\bibinfo  {journal} {Phys. Rev. D}\ }\textbf {\bibinfo {volume} {78}},\ \bibinfo {pages} {103005} (\bibinfo {year} {2008})},\ \Eprint {http://arxiv.org/abs/0807.3872} {arXiv:0807.3872 [gr-qc]} \BibitemShut {NoStop}%
\bibitem [{\citenamefont {Akiyama}\ \emph {et~al.}(2019{\natexlab{b}})\citenamefont {Akiyama} \emph {et~al.}}]{EventHorizonTelescope:2019ggy}%
  \BibitemOpen
  \bibfield  {author} {\bibinfo {author} {\bibfnamefont {K.}~\bibnamefont {Akiyama}} \emph {et~al.} (\bibinfo {collaboration} {Event Horizon Telescope}),\ }\href {\doibase 10.3847/2041-8213/ab1141} {\bibfield  {journal} {\bibinfo  {journal} {Astrophys. J. Lett.}\ }\textbf {\bibinfo {volume} {875}},\ \bibinfo {pages} {L6} (\bibinfo {year} {2019}{\natexlab{b}})},\ \Eprint {http://arxiv.org/abs/1906.11243} {arXiv:1906.11243 [astro-ph.GA]} \BibitemShut {NoStop}%
\bibitem [{\citenamefont {{Gillessen}}\ \emph {et~al.}(2017)\citenamefont {{Gillessen}}, \citenamefont {{Plewa}}, \citenamefont {{Eisenhauer}}, \citenamefont {{Sari}}, \citenamefont {{Waisberg}}, \citenamefont {{Habibi}}, \citenamefont {{Pfuhl}}, \citenamefont {{George}}, \citenamefont {{Dexter}}, \citenamefont {{von Fellenberg}}, \citenamefont {{Ott}},\ and\ \citenamefont {{Genzel}}}]{2017ApJ...837...30G}%
  \BibitemOpen
  \bibfield  {author} {\bibinfo {author} {\bibfnamefont {S.}~\bibnamefont {{Gillessen}}}, \bibinfo {author} {\bibfnamefont {P.~M.}\ \bibnamefont {{Plewa}}}, \bibinfo {author} {\bibfnamefont {F.}~\bibnamefont {{Eisenhauer}}}, \bibinfo {author} {\bibfnamefont {R.}~\bibnamefont {{Sari}}}, \bibinfo {author} {\bibfnamefont {I.}~\bibnamefont {{Waisberg}}}, \bibinfo {author} {\bibfnamefont {M.}~\bibnamefont {{Habibi}}}, \bibinfo {author} {\bibfnamefont {O.}~\bibnamefont {{Pfuhl}}}, \bibinfo {author} {\bibfnamefont {E.}~\bibnamefont {{George}}}, \bibinfo {author} {\bibfnamefont {J.}~\bibnamefont {{Dexter}}}, \bibinfo {author} {\bibfnamefont {S.}~\bibnamefont {{von Fellenberg}}}, \bibinfo {author} {\bibfnamefont {T.}~\bibnamefont {{Ott}}}, \ and\ \bibinfo {author} {\bibfnamefont {R.}~\bibnamefont {{Genzel}}},\ }\href {\doibase 10.3847/1538-4357/aa5c41} {\bibfield  {journal} {\bibinfo  {journal} {\apj}\ }\textbf {\bibinfo {volume} {837}},\ \bibinfo {eid} {30} (\bibinfo {year} {2017})},\ \Eprint
  {http://arxiv.org/abs/1611.09144} {arXiv:1611.09144 [astro-ph.GA]} \BibitemShut {NoStop}%
\end{thebibliography}%


\end{document}